\documentclass[a4paper, 12pt]{article}

\usepackage{graphicx}
\usepackage[utf8]{inputenc}
\usepackage{lineno}
\usepackage{blindtext}
\usepackage{url}
\usepackage{amsmath}
\usepackage{amsthm}
\usepackage{amssymb} 
\usepackage{float, placeins}
\usepackage{caption, subcaption}
\usepackage[font={it}]{caption}
\usepackage[margin=2.0cm]{geometry}
\usepackage{cancel}
\usepackage{xfrac}
\usepackage{hyperref}
\usepackage{booktabs}
\usepackage{siunitx}
\usepackage{accents}
\usepackage{titling}
\usepackage[super]{nth}
\usepackage[toc,page]{appendix}
\usepackage{algorithm}
\usepackage{algpseudocode}
\usepackage{bm} 
\usepackage{arydshln}
\usepackage{tikz}
\usepackage{tikz-3dplot}
\usepackage{datetime2}
\usepackage{authblk}
\usepackage{blindtext}

\newcommand{\sfVeryTiny}{0.25}
\newcommand{\sfTiny}{0.33}
\newcommand{\sfSmall}{0.49}

\usepackage{graphicx}
\usepackage{subcaption}
\usepackage{makecell}

\begin{document}

\date{\today}
\title{Cosmic rays for imaging cultural heritage objects}

\author[1]{Andrea Giammanco\footnote{Corresponding author; andrea.giammanco@cern.ch}}
\author[1]{Marwa Al Moussawi}
\author[2]{Matthieu Boone}
\author[3]{Tim De Kock}
\author[4]{Judy De Roy}
\author[4]{Sam Huysmans}
\author[1]{Vishal Kumar}
\author[1]{Maxime Lagrange}
\author[5]{Michael Tytgat}

\affil[1]{Centre for Cosmology, Particle Physics and Phenomenology (CP3), Université catholique de Louvain, Chemin du Cyclotron 2, B-1348 Louvain la Neuve, Belgium}
\affil[2]{Ghent University Centre for X-ray Tomography, University of Ghent,      Proeftuinstraat 86, B-9000 Gent, Belgium}
\affil[3]{Antwerp Cultural Heritage Sciences (ARCHES), University of Antwerp, Mutsaardstraat 31, B-2000 Antwerp, Belgium}
\affil[4]{Stone Sculpture Studio, Conservation \& Restoration, Royal Institute for Cultural Heritage (KIK-IRPA), Jubelpark 1, B-1000 Brussels, Belgium}
\affil[5]{Elementary Particle Physics Group, Vrije Universiteit Brussel, Pleinlaan 2, B-1050 Brussel, Belgium}

\maketitle

\abstract{In cultural heritage conservation, it is increasingly common to rely on non-destructive imaging methods based on the absorption or scattering of photons ($X$ or $\gamma$ rays) or neutrons. However, physical and practical issues limit these techniques: their penetration depth may be insufficient for large and dense objects, they require transporting the objects of interest to dedicated laboratories, artificial radiation is hazardous and may induce activation in the material under study. 
Muons are elementary particles abundantly and freely produced in cosmic-ray interactions in the atmosphere. 
Their absorption and scattering in matter are characteristically dependent on the density and elemental composition of the material that they traverse, which offers the possibility of exploiting them for sub-surface remote imaging. 
This novel technique, nicknamed “muography”, has been applied in use cases ranging from geophysics to archaeology to nuclear safety, but it has been so far under-explored for a vast category of cultural heritage objects that are relatively large (from decimeters to human size) and dense (stone, metals). 
The development of portable muon detectors makes muography particularly competitive in cases where the items to be analysed are not transportable, or set up in a confined environment. 
This document reviews the relevant literature, presents some exemplary use cases, and critically assesses the strengths and weaknesses of muography in this context. 
}

\begin{center}
	\hrule
\end{center}
 \tableofcontents
\begin{center}
	\hrule
\end{center}


\section{Introduction}
\label{sec:intro}

In the past decades, imaging techniques based on X-rays have been extensively used in the investigation and analysis of our cultural heritage, for their penetrating power in materials. Examples hereof are ubiquitous~\cite{morigi2010application}: imaging sculptures/objects to visualize their composition (different volumes, joints, old restorations, inner structures or reinforcements); imaging mummies, reliquaries, tombs to visualize their content without manipulating the objects; dating wooden artifacts using dendrochronology; etc. However, there are important limitations to X-ray imaging. When looking at large objects, or objects composed of dense materials (compact stone, metal, etc.), the penetrating power of standard X-rays is insufficient. 

By using alternative types of radiation, such as MeV-range X-rays and neutrons, this problem can be partly overcome. 
However, moving valuable objects to an imaging facility is not always feasible due to their size, weight, nature, poor state of preservation or simply the risk of damage during transport. These limitations affect both movable and immovable heritage such as monuments, monumental statues and building decorations, and individual sculptured objects. 
Nevertheless, imaging is key for a better understanding of their composition and can provide important information about their construction techniques, earlier interventions (gluing, reinforcements, additions, etc.), their condition and interrelationships with the environment. Portable and compact set-ups for X-ray fluorescence analysis (XRF) are available for use in cultural heritage studies~\cite{bezur2020handheld,ridolfi2012portable}, where fluorescent X-rays are used to reveal chemical composition, but are intrinsically limited to shallow depths by the high absorption rate of X-rays.  
Portable X-ray computed tomography (CT) systems have been deployed for the in-situ study of art objects~\cite{Albertin2019}. However, these are less performant than fixed laboratory systems and require safety measures during the data-taking campaign. 
Portable fast-neutron sources are also in use~\cite{kerr2022neutron}, and neutron absorption reaches larger depths than X-rays~\cite{kardjilov2018advances}. 
A portable proton accelerator has been recently developed~\cite{taccetti2023machina}, for in situ X-rays (PIXE, Particle Induced X-rays Emission) and low-energy $\gamma$-rays (PIGE, Particle Induced Gamma-rays Emission) detection. 
All these methods suffer the practical issue that ionizing radiation (including X-rays, $\gamma$-rays, protons and neutrons) is strictly regulated because of radiation hazard concerns, and neutrons can also cause the activation of material (i.e. creating radioactive nuclei inside the object under investigation), which is highly undesirable.

\begin{figure}[H]
\centering
\begin{minipage}{14pc}
\centering
\includegraphics[width=12pc]{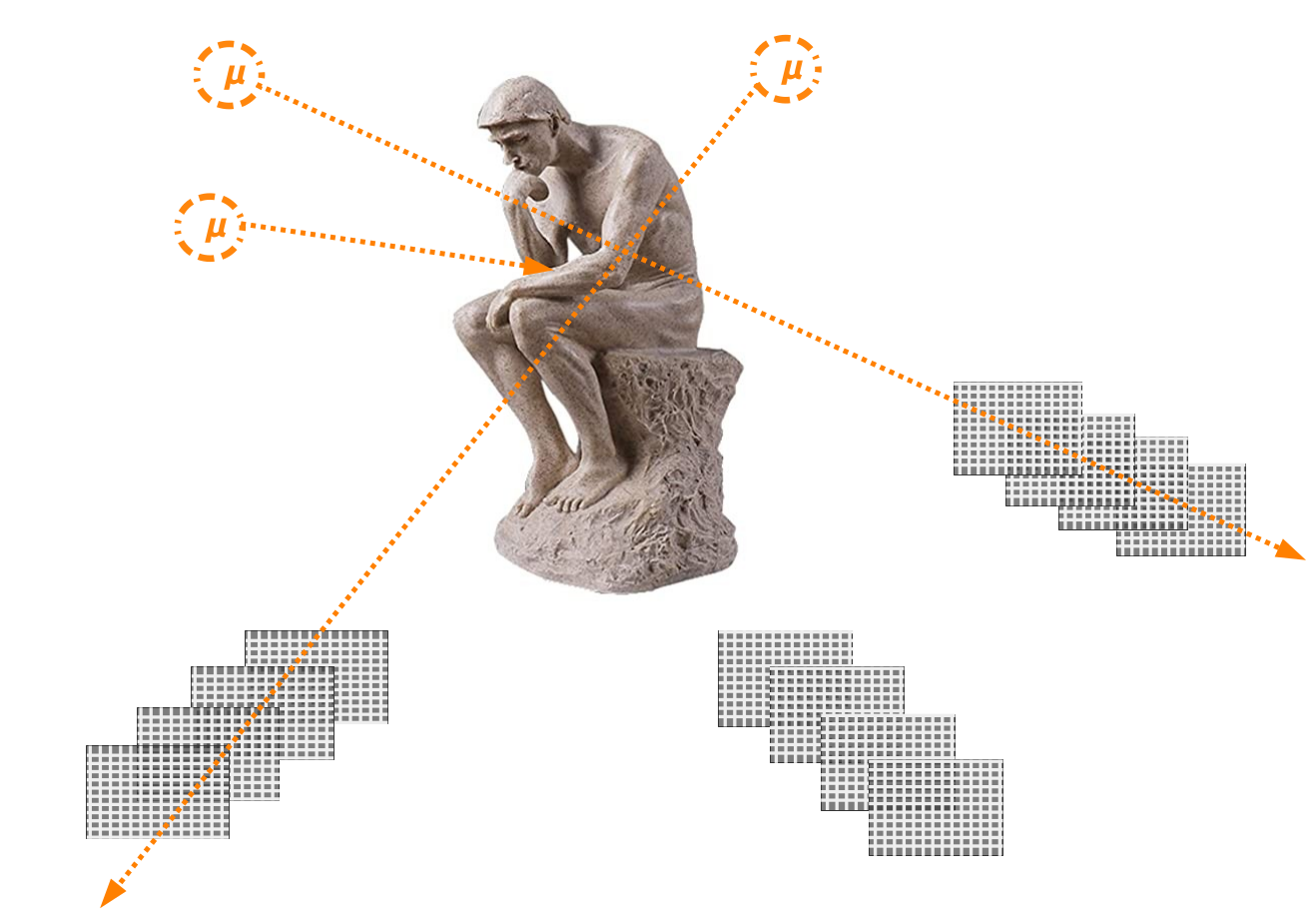}
\caption{In absorption muography, muon trackers are downstream of the object of interest; 3D imaging can be obtained by combining multiple viewpoints.}
\label{fig:sketch-abs}
\end{minipage}\hspace{5pc}
\begin{minipage}{14pc}
\centering
\includegraphics[width=12pc]{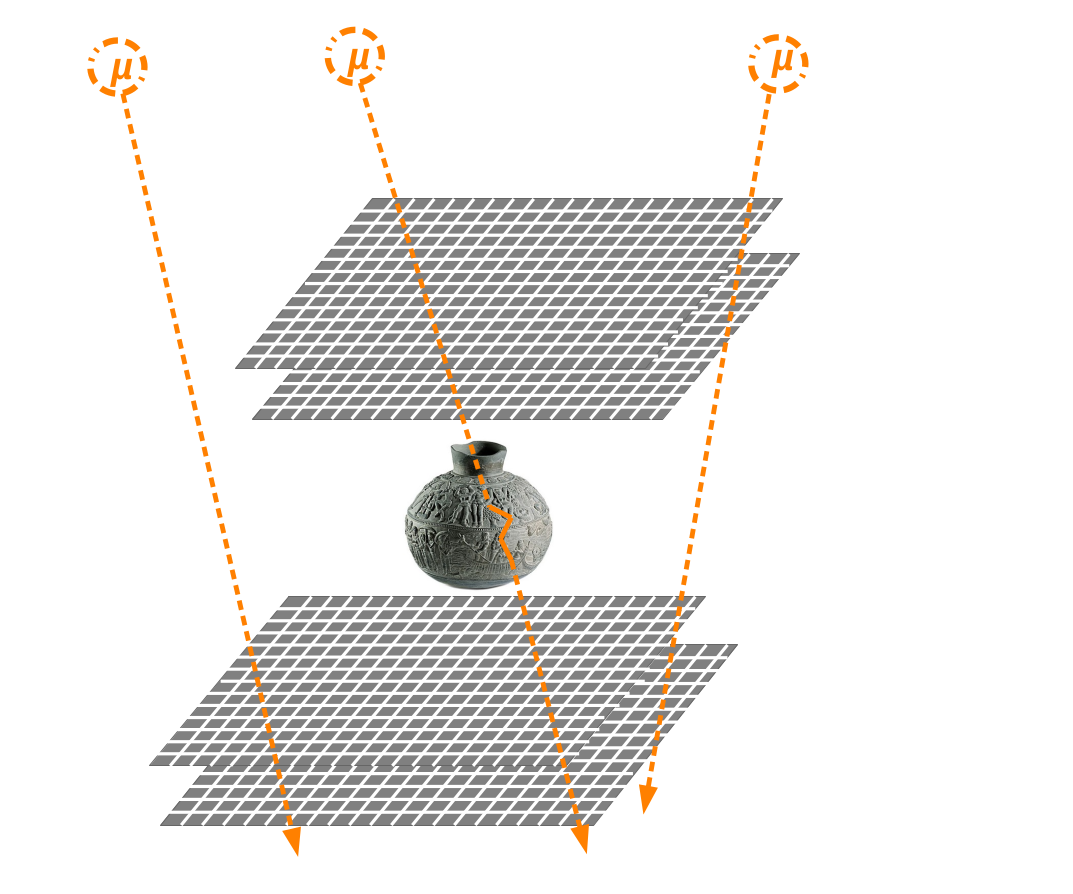}
\caption{In scattering muography, the object under investigation is ”sandwiched” between muon trackers.} 
\label{fig:sketch-scatter}
\end{minipage} 
\end{figure}

In this paper we argue for portable and safe imaging based on a different elementary particle, the muon ($\mu$). 
Mostly produced at high energy (most of them at order GeV), muons easily penetrate through large depths; they are harmless and available everywhere. 
These characteristics inspired several practical applications in sub-surface imaging, collectively known under the umbrella term of ``muography''~\cite{Bonechi:2019ckl,IAEA2022,MuographyBook}. 
Two main muography techniques exist, to which in this paper we refer as absorption-based and scattering-based. 

Absorption(-based) muography exploits the loss of energy (mostly by ionization) of muons when passing through any material; when its initial kinetic energy is completely exhausted, the muon stops and decays into an electron\footnote{To be precise, negatively charged muons decay into electrons while positively charged muons, also called anti-muons, decay into positrons, also called anti-electrons. For sake of brevity, and following the conventions of high-energy physics, through this paper we employ the word electron to also indicate positrons, and with the word muon we do not make distinction between the two charge states.} and two neutrinos, which are not detected.
Absorption muography is directly sensitive to the matter density of the material, which gives the possibility to discriminate between different hypotheses on the inner composition of a large object, and also assess specific conditions such as the moisture content. 
Similar to conventional radiography, absorption muography naturally yields 2D images. In this technique, the trajectory of the muon must be recorded by a tracker, which measures its zenith ($\theta$) and azimuth ($\phi$) angles. 
The tracker is usually an array of position-sensitive layers of particle detectors, although other tracker types exist~\cite{Bonechi:2019ckl}.
Subsequently, the 2D distribution of the muon rate in the $\theta-\phi$ plane is compared between data sets with and without the object of interest in front of the tracker, to get a transmission map, which in turn can be translated into density projections or opacity maps (where the opacity is defined as the integral of density over a line of sight). 
To obtain 3D images, 2D density projections observed from different points of view can be combined through inversion algorithms. 
A possible arrangement of muon trackers around a statue, for 3D imaging, is sketched in Fig.~\ref{fig:sketch-abs}. 

Scattering(-based) muography, first proposed in 2003~\cite{Borozdin2003}, instead exploits the diffusion of muons due to the electric charge of the atomic nuclei in matter. The average angular deviation of the muon trajectory has a very strong dependence on the atomic number of the substance traversed, i.e. the number of protons in the nucleus of its atoms, which means that this method can be used for elemental discrimination. 
In practice, the muon trajectory must be reconstructed both upstream and downstream of the volume of interest in order to measure the angular deviation. As a consequence, the object under investigation must be "sandwiched" between muon trackers, as illustrated in Fig.~\ref{fig:sketch-scatter}, which becomes impractical or expensive when the object is large. 
Scattering muography is an intrinsically 3D imaging method, as it is possible to reconstruct the muon trajectory in 3D space. In the most popular analysis method for scattering muography, the most likely position of the largest scattering event is estimated for each muon by extrapolating the incoming and outgoing trajectories measured by the two trackers and finding their point of closest approach (POCA). From the frequency of POCA points attributed to each voxel, and the average scattering angle in each, it is possible to test hypotheses on the identity of the material voxel by voxel. This POCA method implicitly assumes that only one large-scattering event happened and that all other scattering events can be neglected; although this approximation is satisfactory for relatively small objects, several other algorithms have been developed in the specific literature that do not rely on this assumption~\cite{barnes2023cosmic}.

Each of the two muography techniques has its own sensitivity, applicability, and limits. 
In absorption muography, a single muon tracker is able to measure the 2D projection of matter density, and the combination of measurements from different viewpoints can give a 3D density map. 
However, it provides no material discrimination apart from density, and small-size or low-density objects do not stop enough muons to provide sufficient contrast. 
In scattering muography, at least two muon trackers are needed, upstream and downstream of the object of interest, to reconstruct the muon trajectory before and after passing through it. 
This method naturally yields 3D information, and is sensitive to elemental composition because the width of the scattering angle distribution is a function of atomic number Z. 
However, it is
impractical for human-sized statues, as the object of interest must fit between the two trackers.
Either the object of interest is moved inside the set-up, or a rather complex installation of the detectors must be performed around the object. 
Therefore, this method is appropriate for relatively small objects. 

Cosmic muons pose no hazard, being part of the natural and ubiquitous background that constantly crosses our bodies and the objects of art. Muon detectors can be made relatively inexpensive and transportable. Muography images are generally easy to interpret. Moreover, 3D inversion of muography data is conceptually simpler than many remote sensing technique. 
But muography also has some fundamental limitations:
the muon flux is relatively small, O(100 Hz/m$^2$); the scattering in matter blurs the images, and the amount of blurring depends on the muon energy, whose spectrum is very model-dependent and which is hard to measure per-particle without making the size and cost of the detectors increase drastically. 
These limitations may be overcome with dedicated developments, as elaborated in this paper.


This paper is structured as follows: 
Section~\ref{sec:state-of-the-art-ch} summarizes the state of the art in non-destructive quantitative analysis methods used in cultural heritage preservation that fulfill the request for portability, while Section~\ref{sec:state-of-the-art-muo} presents examples of prior applications of muography to (large-scale) cultural heritage. 
In Section~\ref{sec:examples}, we present a simulation study whose aim is to identify in which regimes the two main muography methods are most promising, and which limitations must be overcome in future studies. 
Section~\ref{sec:future-studies} elaborates on a few future case studies of great interest for cultural heritage preservation.
Finally, in Section~\ref{sec:conclusions} we draw the conclusions.
Appendix~\ref{sec:appendix} provides additional explanations on the algorithms employed in this first study.




\section{State of the art in portable non-destructive test methods for cultural heritage}
\label{sec:state-of-the-art-ch}


A condition assessment is indispensable for the conservation of cultural heritage objects. Whether it’s a piece of art, a monument or another object with cultural value, this starts with a profound visual observation of the object, supported by a detailed photographic recording of the object to document the state of preservation at that time. Surface 3D-models can be achieved easily, apart from 3D scanning techniques, by combining different views from different angles from the object, a technique known as photogrammetry. These 2D or 3D images serve as the base for detailed mappings of observed materials, surface treatments and pathology. Additional portable methods are available for more profound diagnostics. 

\textit{Metal detector:} Non-ferrous and ferrous metal elements, like inner supporting structures in sculptures, can be detected by a portable metal detector (stud or wall detector), designed for the detection of wires, cables and pipes in buildings. It is a simple handheld device that sends an electric signal that interacts with the density and composition of an object in a building material, a sculpture or an architectural decoration. These electronic detectors use sensors to detect any changes in the dielectric constant of the support. The dielectric constant will change when the scanner sensor passes over the stud. 
This device is often limited to a maximal depth of 12 to 15 cm for ferrous metals and even less for non-ferrous metals. It is possible to locate the center of the metal element but no information can be obtained about the dimensions and the exact orientation of the metal elements. In the case of sculptures, the use of the detector can be obstructed by the complex form or composition of the sculpture itself.

\textit{Radar:} Imaging radar technologies, in particular ground penetrating radar (GPR) detect subsurface dielectric discontinuities through the reflection of emitted electromagnetic waves. Originally developed for geophysical surveying, applications in cultural heritage are mainly oriented towards structural investigations.

\textit{IR-thermography:} Infrared thermography measures the infrared (thermal) radiation coming from the material under investigation and renders the image of the surface area in colours or in grey scale, in relation to a temperature scale for an assumed material emissivity ~\cite{avdelidis2004applications}. It can be used to spot surface anomalies, but it is often difficult to interpret observations without the support of optical imaging or other techniques. It has been used to detect near-surface differences in moisture conditions, as evaporation of moisture chills the surface relative to drier areas.   

\textit{Ultrasonic Pulse Velocity (UPV)} measurement is a technique that measures the travel time of an ultrasonic pulse (typically a P-wave), sent by a transmitter and captured at a known distance from the transmission source. The UPV is the ratio between the path length (the distance between transmitter and receiver) and travel time. Anomalies such as fissures, cracks, cavities and decohesion reduce the UPV relative to sound material, allowing for their indirect assessment~\cite{Weiss2002}. Some standard materials, like Carrara marble, have well characterized UPV values supported by multiple studies, while others need benchmarking case by case, putting limits on its usability. The distance between the transmitter and receiver is dependent on the wave frequency, but typically requires $> 10$ cm. In case of a severely damaged sculpture (for example the Egyptian Khonsou sculpture, Sec.~\ref{sec:inner-cracks}) it is impossible to interpret these values in order to determine the pattern of the cracks. Some authors have argued other parameters, such as spatial attenuation, to be more reliable~\cite{Martinez-Martinez2011}. However, that would require the measurement setup to be equipped with a (more expensive) oscilloscope, which is not always the case.

\textit{Ultrasonic tomography} is based on a series of UPV measurements. With this technique it is possible to define areas in objects where fractures and cracks are present. Ultrasonic tomography is mostly used in concrete industries but some studies have been performed on cultural heritage~\cite{Secanellas2022,Ruedrich2011}. This technique is often used on two dimensional views at a certain depth of the object but also three dimensional analyses are possible. 
3D scanning of the surface increases the potential use of ultrasonic tomography, as this can serve as a base to determine the location of transmitter and receiver as it simplifies the identification of the x, y and z positions of the transmitter and receiver of the ultrasonic waves on the statue. Nevertheless, these measurements are labour intensive and become more complex with increasing complexity of the object's shape.  As measurements must be taken on the surface of the object under investigation, this must be fully accessible along all sides.


\section{State of the art in muography}
\label{sec:state-of-the-art-muo}


The first application of muography dates back to the middle of the past century~\cite{George1955}, to measure the overburden of a tunnel. It then took 15 years to see the second, and arguably most famous, application of the same principle, when L. Alvarez’s team used a spark chamber as a muon tracker to search for possible hidden chambers inside an Egyptian pyramid~\cite{Alvarez1970}. More applications followed, especially in the 21st century, and muography is nowadays a booming research area, with a steadily growing trend of publications~\cite{Holma_Joutsenvaara_Kuusiniemi_2022}. 
The interested reader can refer e.g. to the reviews in \cite{Bonechi:2019ckl} and \cite{IAEA2022}.
Several ongoing efforts in absorption-based muography focus on targets that are very large, i.e. from tens of meters (e.g. buildings - both ancient and modern - and civil infrastructure such as bridges) to kilometres (mountains, and in particular volcanoes)~\cite{MuographyBook}, while after the invention of scattering-based muography in 2003~\cite{Borozdin2003}, the latter technique has been applied or advocated for many uses in the nuclear safety sector (e.g. nuclear waste assays, nuclear smuggling prevention at border controls, verification of non-proliferation treaties), exploiting its unique sensitivity to the number of protons in nuclei. 
Because of the need to "sandwich" the volume of interest between upstream and downstream detectors, scattering muography is generally limited to smaller objects than absorption muography.

Muography has already been successfully used for studies of large-scale cultural heritage. 
One of the most renowned examples is the ScanPyramids project, based on three different detector technologies (plastic scintillator detectors, micro-pattern gaseous detectors, nuclear emulsions), which in 2017 announced an unexpected low-density anomaly in Khufu’s Great Pyramid~\cite{Morishima2017}. Additional data have been collected since then, leading to the recent announcement of a corridor-like structure of about 9 m length with a transverse section of about 2 m by 2 m~\cite{procureur2023precise}. Confidence in the solidity of these conclusions was given by the addition, since 2020, of measurements with GPR and ultrasonic testing~\cite{elkarmoty2023localization}. This motivated finally the visual inspection via an endoscope, which gave uncontroversial confirmation of the claim.

In another example, density anomalies (potentially posing safety hazards) have been found in a rampart of a defensive wall of Xi’an (China)~\cite{Xian-walls}. 
In the Svyato-Troitsky Danilov monastery (Russia),
areas of significantly higher density within immured vaults (interpreted as walls and partition walls) and unknown voids (possible ancient crypts or air ducts) have been revealed through the use of muon detectors based on nuclear emulsions~\cite{alexandrov2022noninvasive}.
Nuclear emulsions have also been exploited in a muography survey of an archaeological site located in a highly populated district of Naples (Italy), ten meters below the current street level, which observed some unknown structures, include one compatible with the hypothesis of a hidden, currently inaccessible, burial chamber~\cite{tioukov2023hidden}.

The examples above all made use of absorption muography, and in general many methodological developments from geophysical applications (see~\cite{MuographyBook,lechmann2021muon} for dedicated reviews) can also be applied to the study of very large human-made objects. 
However, a proposal has been made to search for iron chains within masonry in the dome of the Florence cathedral (Italy)~\cite{Guardincerri2018} using scattering-based muography, and a proof-of-principle test has been carried out on a mock-up wall, demonstrating the conceptual validity of the method. 
Another method has been invented more recently, alignment-based muography, which exploits the fact that, in absence of electric or magnetic fields, the muon trajectories are straight lines~\cite{bodini2007cosmic}. This method requires two muon telescopes, positioned respectively on a fixed reference system (e.g. a structural element of a building) and on the point whose movements need to be monitored. If the actual position of the latter is not where the track reconstruction algorithm expects it to be, the muon’s half-trajectories do not meet, and the degree of misalignment can be monitored as a function of time as part of an alarm system for the stability of the building.
A proposal has been made to employ this method for the long-term monitoring of the stability of Palazzo della Loggia in Brescia (Italy)~\cite{donzella2014stability}.


In all these examples the particle detectors are relatively large, and therefore not easily transportable. This is due to the large size of the objects of interest, which implies a large attenuation of the flux of muons passing through it, which demands a large cross-sectional area (order m$^2$) in order to maximize the usable statistics within a reasonable data-collection time. 
Instead, in the use cases outlined in Section~\ref{sec:future-studies} we argue for the development and deployment of compact and portable particle detectors. 
Given the confined space available in e.g. a museum or a church, they must be small.
Additionally, researchers may be granted access only outside of visiting hours, therefore the detectors must easily deployable and removable. 
And when we need to resolve relatively small features within the object, as in the examples of Sections~\ref{sec:inner-features} and~\ref{sec:inner-cracks}, spatial resolution of the detectors is an important factor. 

A few teams are developing portable muography detectors. For usage underground (e.g. mining exploration), very specialized detectors are developed that fit inside boreholes (first proposed in \cite{Malmqvist1979}; see e.g. \cite{cimmino2021new} for a recent example); we do not elaborate on those, as the case studies proposed in Section~\ref{sec:future-studies} are not underground. 
R\&D projects of portable detectors that would be in principle suitable for the applications that we propose include, among others: MIMA~\cite{Baccani:2018nrn}, based on scintillating bars coupled with silicon photomultipliers; MUST2~\cite{Roche_2020}, a time-projection chamber (TPC) based on micropattern gaseous detectors called MicroMegas; a project at the University of Kyushu~\cite{Kyushu2018} based on scintillating fibres; a project at Wigner RCP~\cite{hamar2022underground} based on multi-wire gaseous detectors; and our own detector project based on Resistive Plate Chambers, whose various development steps have been detailed in a few proceedings~\cite{Wuyckens2018,Basnet2020,Basnet:2022cds,Gamage_2022,Gamage:2022qiz,Moussawi2021,Kumar:2023irj}.

\section{Simulation studies}
\label{sec:examples}

This paper advocates for the adoption of portable and safe muography as a promising imaging approach for cultural heritage studies in a regime that is new for muography (relatively low size) while being beyond reach for methods based on other radiation sources. We aim at demonstrating that both scattering and absorption muography techniques can play a role in cultural heritage imaging as they are complementary in terms of detection apparatus cost, imaging capabilities and target object size. 
To illustrate the potential applications and limitations of muography, we report in this section a preliminary imaging study of statues of various sizes based on simulated muon tomography data.

\begin{figure}[H]
\centering
\begin{minipage}{.35\linewidth}
\centering
\includegraphics[width=.4\linewidth]{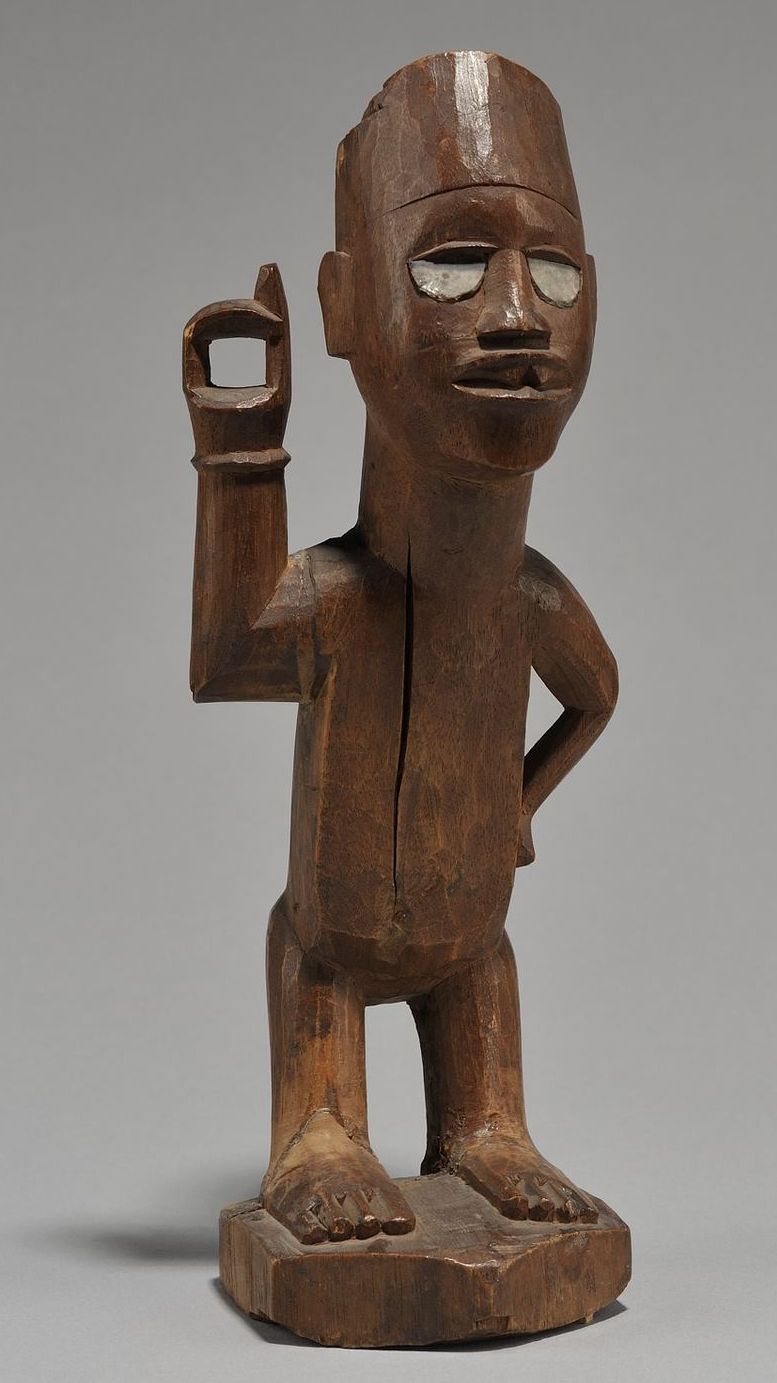}
\caption{Picture of the wooden statue at the AfricaMuseum of Tervuren that we used as model. From project ToCoWo (\url{https://tocowo.ugent.be/}).}
\label{fig:african-statue}
\end{minipage}\hspace{2pc}
\begin{minipage}{.5\linewidth}
\centering
\includegraphics[width=.9\linewidth]{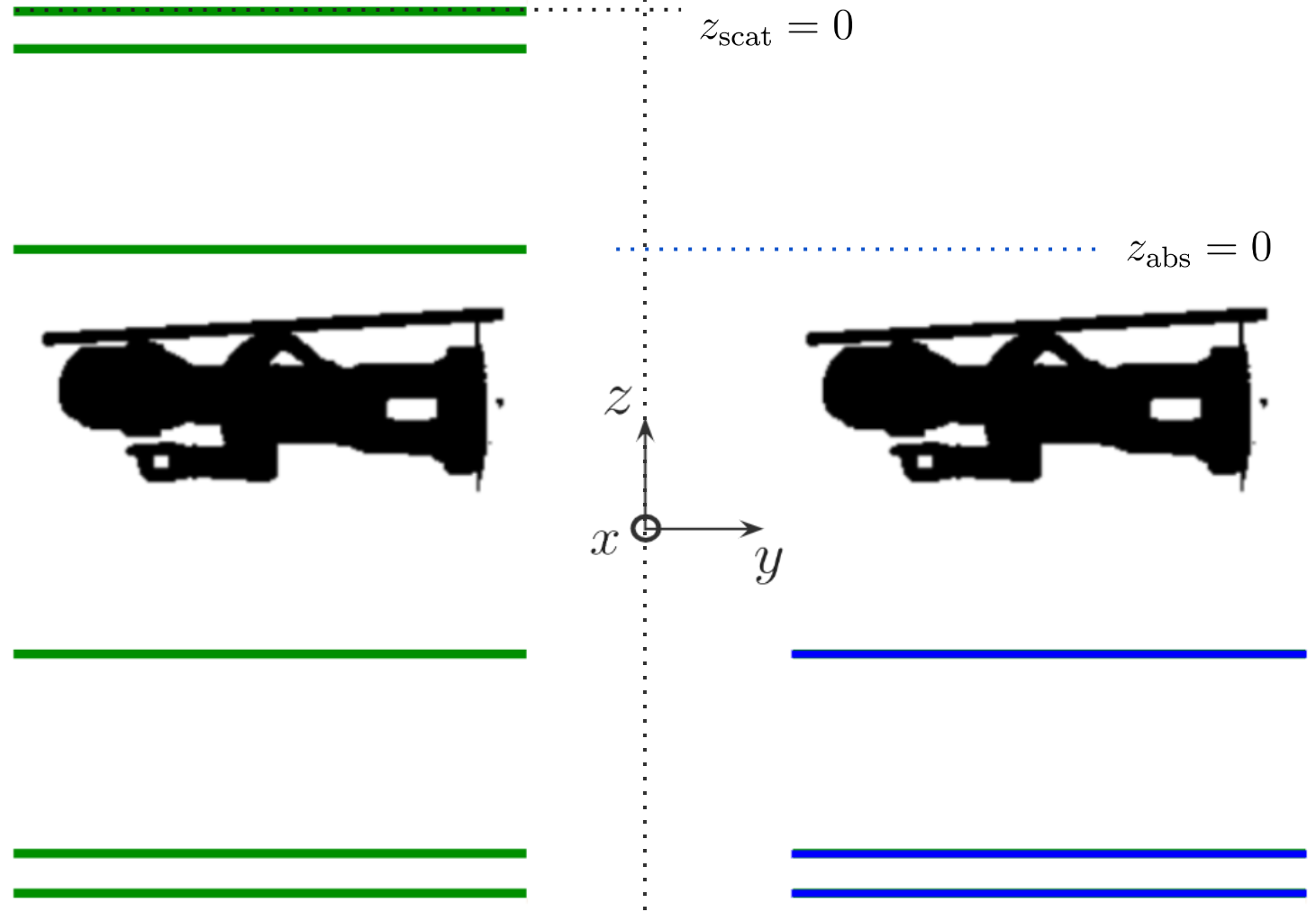}
\caption{Geant4 simulation setup; green and blue panels represent the scattering and absorption setups, respectively.}
\label{fig:setup}
\end{minipage} 
\end{figure}


All data used in this study have been generated with the GEANT4~\cite{GEANT4} software for the simulation of the passage of particle through matter, interfaced to the Cosmic-ray Shower Library (CRY~\cite{CRY}) in order to simulate a realistic cosmic muon source. The geometrical model of an African statue (Fig.~\ref{fig:african-statue}) was implemented in GEANT4 as a mesh file which allowed to preserve the high granularity of the original model obtained from a CT scan. Although the actual statue is made of wood, in this study the material was changed to marble with a density of $2.71~{\rm g/cm}^{3}$. 
To investigate the potential of muography for material identification, we simulate the insertion of hidden cylinders of bronze within its internal structure. Three different size regimes have been studied, as summarized in Table~\ref{tab:Simulation}, where the simulated statue's size is the same as the original statue or scaled by factors two or four, respectively.

In this first exploratory study, we model an ideal detection apparatus (i.e. with 100\% efficiency and perfect resolution) made up of six or three planes, for scattering or absorption muography respectively, as illustrated in Fig.~\ref{fig:setup}. The amount of simulated muons corresponds to eight hours of continuous data taking.

The following sections will show the imaging performance of scattering and absorption muography using various image reconstruction, clustering and image processing algorithms. 



\begin{table}[ht]
\centering
\begin{tabular}{|c|c|c|c|}
\hline
\bf Scenario & \bf \makecell{Statue \\ size [cm$^3$]}  & \bf \makecell{Cylinder\\ radius [cm]}  \\ 
\hline

I & $80 \times 30 \times 30$ & 5  \\ \hline
II & $160 \times 60 \times 60$ & 10  \\ \hline
III & $320 \times 120 \times 120$ & 20 \\ \hline 
\end{tabular}
\caption{Summary of the different simulated scenarios.}
\label{tab:Simulation}
\end{table}

\subsection{Scenario I: small size regime}
\label{sec:scenario1}

The statue's size simulated in scenario is  $80 \times 30 \times 30 \:\text{cm}^{3}$, which is quite small by muography standards, making the scattering method the most relevant. This section presents the imaging capabilities of various scattering tomography algorithms complemented by algorithms relevant for material identification.

As described in Section~\ref{sec:intro}, scattering muography is based on the measurement of muon deflections through the object. The scattering angle is measured by extrapolating the incoming and outgoing trajectories measured by the two sets of detectors and used by the reconstruction algorithms to infer the relative density of the target volume. While a plethora of algorithms are available in the literature, this first section will rely on the Point Of Closest Approach (POCA)\cite{POCA}, which is detailed in Appendix~\ref{sec:appendix}.

\begin{figure}[H]
    \centering
    \includegraphics[width = \textwidth]{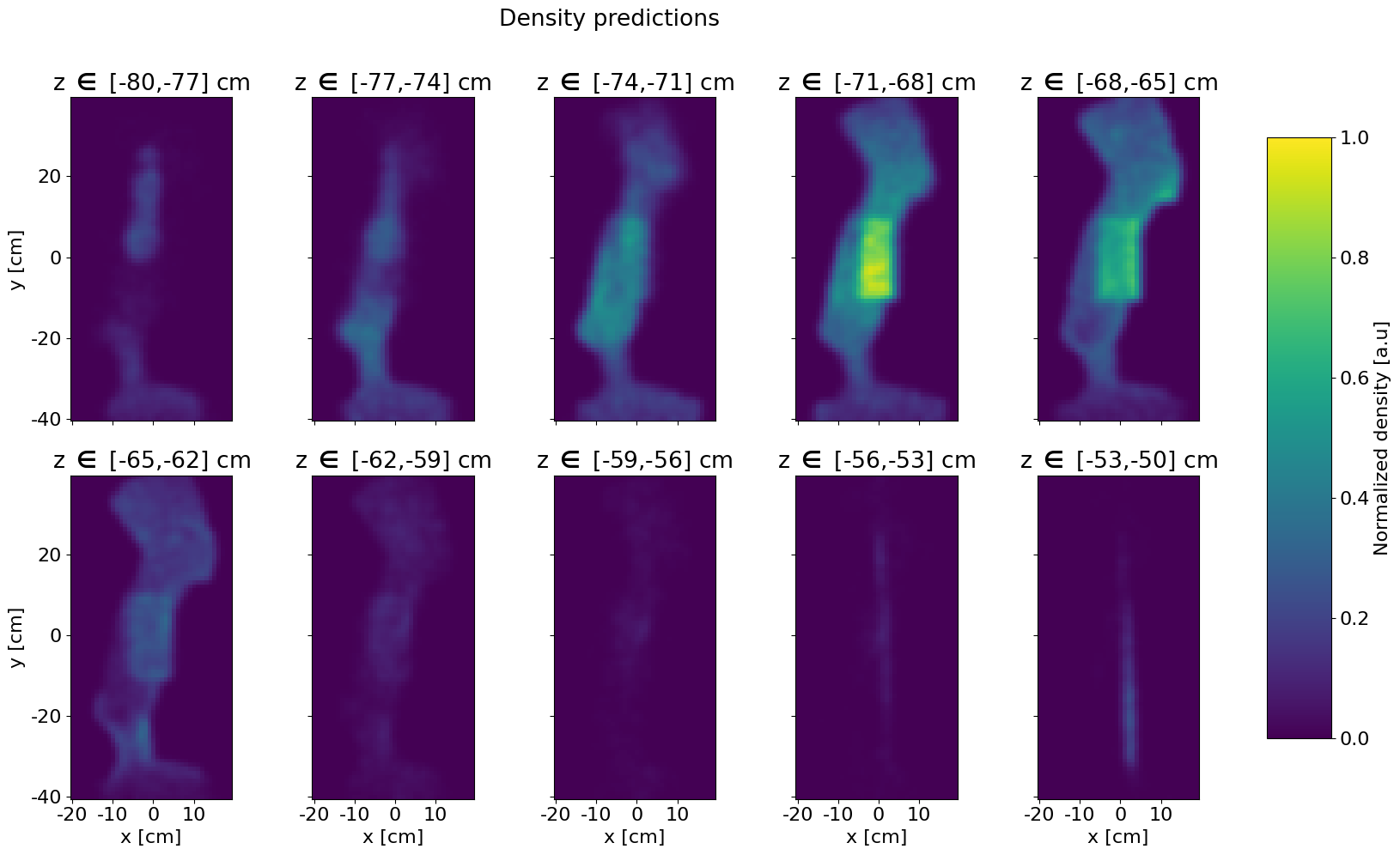}
    \caption{Normalized inferred density computed as number of POCA points per voxel. Density predictions are shown per slice of 3 voxels layers along the vertical direction $z$, with a voxel size of $10 \:\text{mm}$.}
    \label{fig:scale4_scat_poca_per_vox}
\end{figure}

 The volume at study is divided into three-dimensional pixels (voxels) and the number of POCA points per voxel is used as density prediction. In order to compare density predictions across this document, the density predictions have been normalized. Figure~\ref{fig:scale4_scat_poca_per_vox} shows the density predictions by slice of 3 horizontal voxels layers along the $z$ direction. Predictions have been smoothed using a Gaussian kernel smoother, assuming that material density is locally uniform. Such procedure is a standard practice in image processing and is detailed in Appendix~\ref{sec:appendix}. The bronze cylinder can easily be identified in Figure~\ref{fig:scale4_scat_poca_per_vox} since it appears as a higher density region somewhere between $z = -74 \:\text{cm}$ and $z = -62 \:\text{cm}$. The true location of the center of the cylinder is $(x,y,z) = (0,0,-71)~\text{cm}$.

The reconstructed data obtained from the POCA approach consist in a collection of 4-dimensional data points: the $x, y, z$ location of the point as well as the associated scattering angle. In such a context of low-dimensional data, clustering algorithms can shine by gathering points sharing similar proprieties and thus providing a material discriminator. 

\begin{figure}[H]
    \centering
    \includegraphics[width=0.7\textwidth]{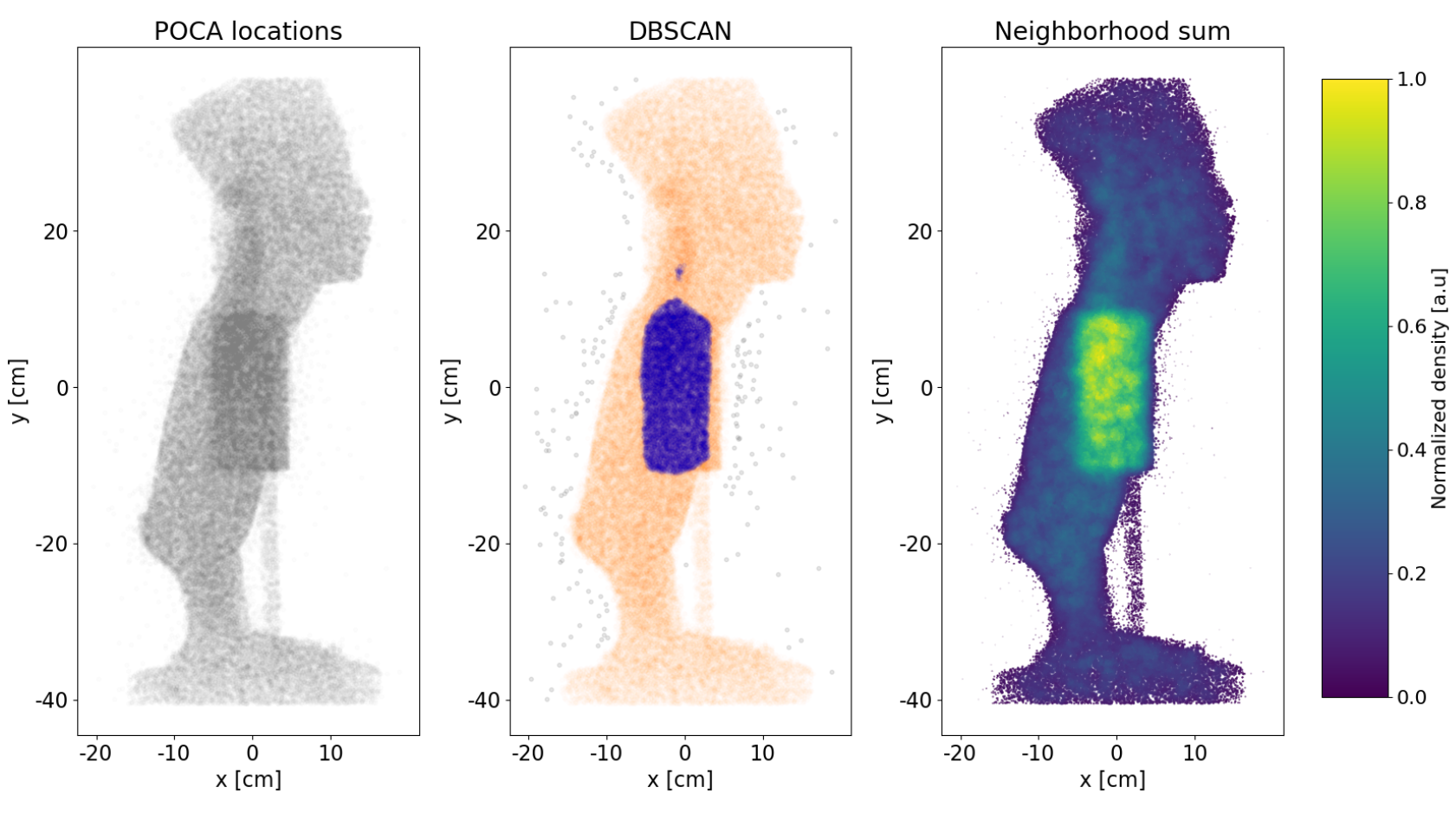}
    \caption{Left: POCA locations. Center: DBSCAN classification based on POCA locations, where marble and bronze are indicated with orange and blue, respectively. Right: neighborhood sum classification based on POCA locations and scattering angles.}
    \label{fig:scale4_scat_clustering}
\end{figure}

The DBSCAN algorithm~\cite{DBScan} discussed in Appendix~\ref{sec:appendix}, is capable of identifying clusters of points within a dataset. These clusters are defined as areas of higher density compared to the rest of the data. In this example, the input for DBSCAN is bidimensional, consisting of the x and y position information of the POCA points (depicted in Figure~\ref{fig:scale4_scat_clustering} left), but the algorithm can be generalized to solve problems in three or higher dimensions. In Figure~\ref{fig:scale4_scat_clustering} center, a nested two-layer DBSCAN is employed. It first effectively separates objects from the noisy background POCA points (grey) and subsequently distinguishes between marble (orange) and bronze (blue) materials.

So far, in Figures~\ref{fig:scale4_scat_poca_per_vox} and~\ref{fig:scale4_scat_clustering} left and center, only the positional information has been utilized for visualization and material discrimination. However, the Neighborhood sum~\cite{kumar2022comparative} offers a unique clustering mechanism by incorporating both position and scattering angle information. The method's details are presented in Appendix~\ref{sec:appendix}. In the current scenario, the x and y position information from the POCA points are utilized to compute the Neighborhood sum, incorporating their respective scattering angles.
The Neighborhood sum information is illustrated in Figure~\ref{fig:scale4_scat_clustering} right, effectively delineating regions of higher density and high scattering angle, thereby distinguishing between marble and bronze.

\subsection{Scenario II: average size regime}
\label{sec:scenario_ii}

A relevant question for planning a muography investigation of an object is whether scattering or absorption is the most appropriate method. Based on which is chosen, the arrangement of detectors is rather different as illustrated in Fig.~\ref{fig:setup}. The scenario II described in Table~\ref{tab:Simulation} is a challenging regime for both absorption and scattering muography. Because the scattering technique requires a set of detection planes above and below the object, the overall price scales faster with the detection area than the absorption technique. The statue's thickness being $60\:\text{cm}$, is rather thin by absorption muography standards as muons loose a small fraction of their kinetic energy making their decay within the object unlikely. Both datasets use similar detection set up made of $1\:\text{m}^2$ detection panels.

\paragraph{Scattering muography}

\begin{figure}
    \centering
    \includegraphics[width=.8\linewidth]{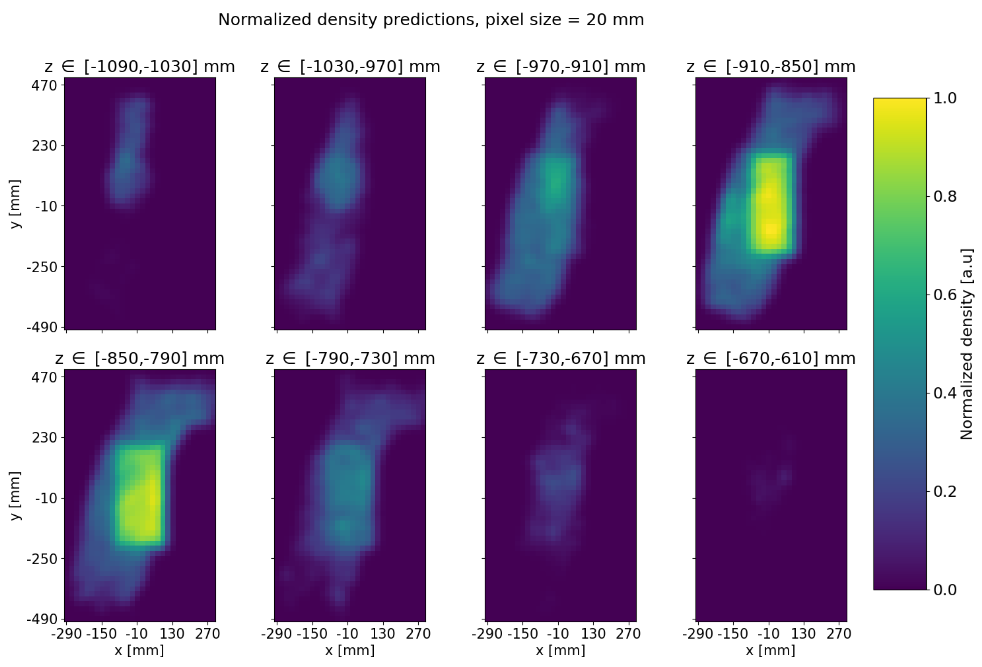}
    \caption{Normalized density predictions for the scattering technique using the Binned Clustered Algorithm. Predictions are shown by slice of 3 voxels layers along z (vertical axis).}
    \label{fig:scale_8_scat_slices}
\end{figure}

Similarly to Section~\ref{sec:scenario1}, density predictions are obtained from the POCA locations. Additionally, the Binned Cluster Algorithm (BCA) was used \cite{BCA}, which incorporates a clustering algorithm utilizing the recorded values of the muons' scattering angle. 
BCA tends to outperform simple POCA when the measured scattering angles associated to the POCA locations are large enough, therefore it is more appealing for this scenario than for the low size regime as the average scattering angles are larger for longer pathlengths.  
To mitigate statistical artifacts, the density predictions are then smoothed using a Gaussian smoother and are presented on Figure ~\ref{fig:scale_8_scat_slices}.

\paragraph{Absorption muography}

\begin{figure}
    \centering
    \includegraphics[width=.8\linewidth]{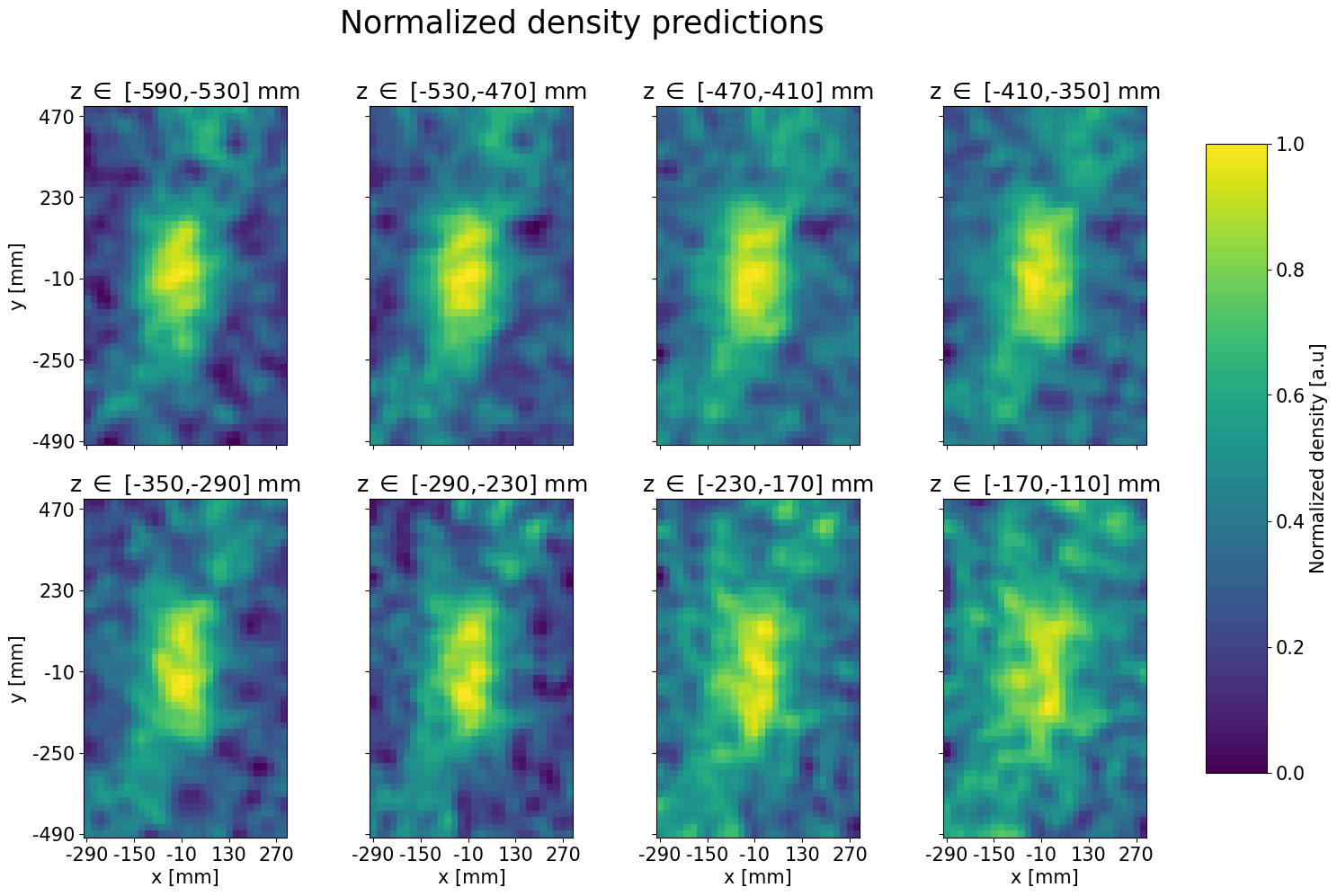}
    \caption{Normalized density predictions for the absorption technique and using the back projection algorithm. Predictions are shown by slice of 3 voxels layers along z (vertical axis).}
    \label{fig:scale_8_abs_slices}
\end{figure}

The voxel-wise density predictions $\rho$ are computed as the ratio between the recorded muon flux through the object and the free sky flux $\rho = \frac{N_{\text{object}}}{N_{\text{freesky}}}$. Both $N_{\text{object}}$ and $N_{\text{freesky}}$ are estimated using a custom back-projection algorithm inspired from \cite{BonechiBP} and \cite{ASR}, which is further described in Appendix~\ref{sec:appendix}. The thickness of the object being rather small compared to absorption muography standards, it is rather unlikely for a muon to decay within the object. In order to compensate for this effect only low energy muons with kinetic energy $E_{\text{kin}} < 1~\text{GeV}$ were selected. Even though the detection setup used does not allow to measure the muons' kinetic energy, a few techniques are available in the literature. Some of them are based on continuous variables such as the track fit's $\chi^2$ that can be used as proxies for energy~\cite{vanini,ANGHEL201512}, while others use a simple energy selection veto based on passing or not a known amount of material downstream of the setup~\cite{sato2022measurement}. In this study we assume that the latter method is used, being cheap and effective. 
Density predictions are also smoothed using the aforementioned Gaussian smoother procedure and are presented on Figure~\ref{fig:scale_8_abs_slices}.

 
\paragraph{Scattering vs absorption}

\begin{figure}
    \centering
    \includegraphics[width=.48\linewidth]{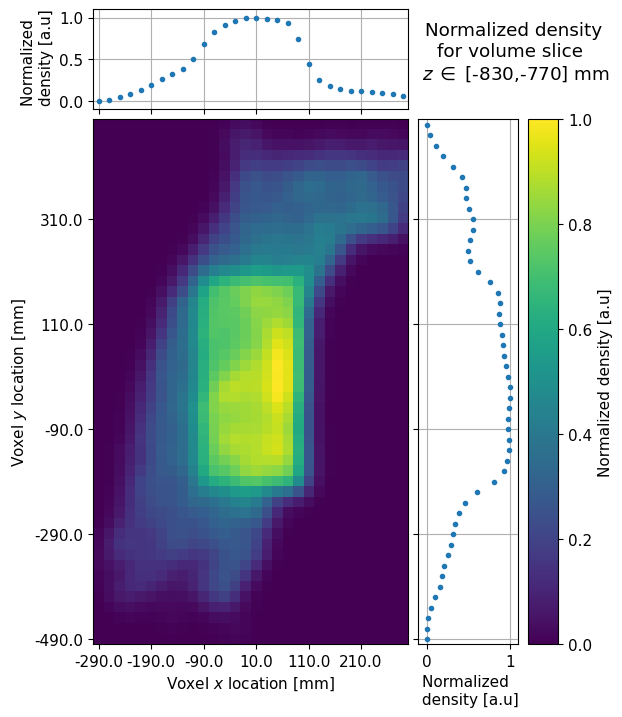}
    \includegraphics[width=.49\linewidth]{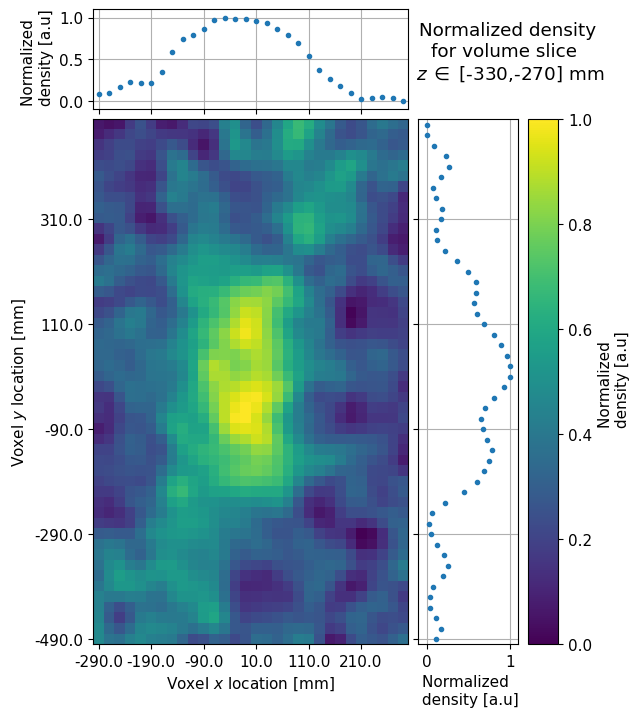}
    \caption{Normalized density predictions for scattering (left) and absorption (right). Predictions are shown by slice of 3 voxels layers along z (vertical axis). The true locations of the 20 cm diameter cylinder are respectively $x, y, z, = (0, 0, -310)~\rm{mm}$ and $x, y, z, = (0, 0, -810)~\rm{mm}$.}
    \label{fig:scale_8_scat_abs_comp}
\end{figure}

Figure~\ref{fig:scale_8_scat_abs_comp} compares scattering and absorption reconstruction for the volume's slice corresponding to the actual position of the bronze cylinder. The presence of a high density material can be clearly identified. Scattering clearly outperforms absorption given that it is sensitive to marble. Absorption predictions are noisier but such effect can be mitigated by increasing the acquisition time.

\subsection{Scenario III: large size regime}

In this regime, absorption becomes more relevant given the large size of the object. Muons lose more kinetic energy while propagating through the thickness of marble, which make then more likely to decay. As a consequence, the measured transmission ratio gets significant enough to provide density estimate of the object.

\begin{figure}
    \centering
    \includegraphics[width=.99\linewidth]{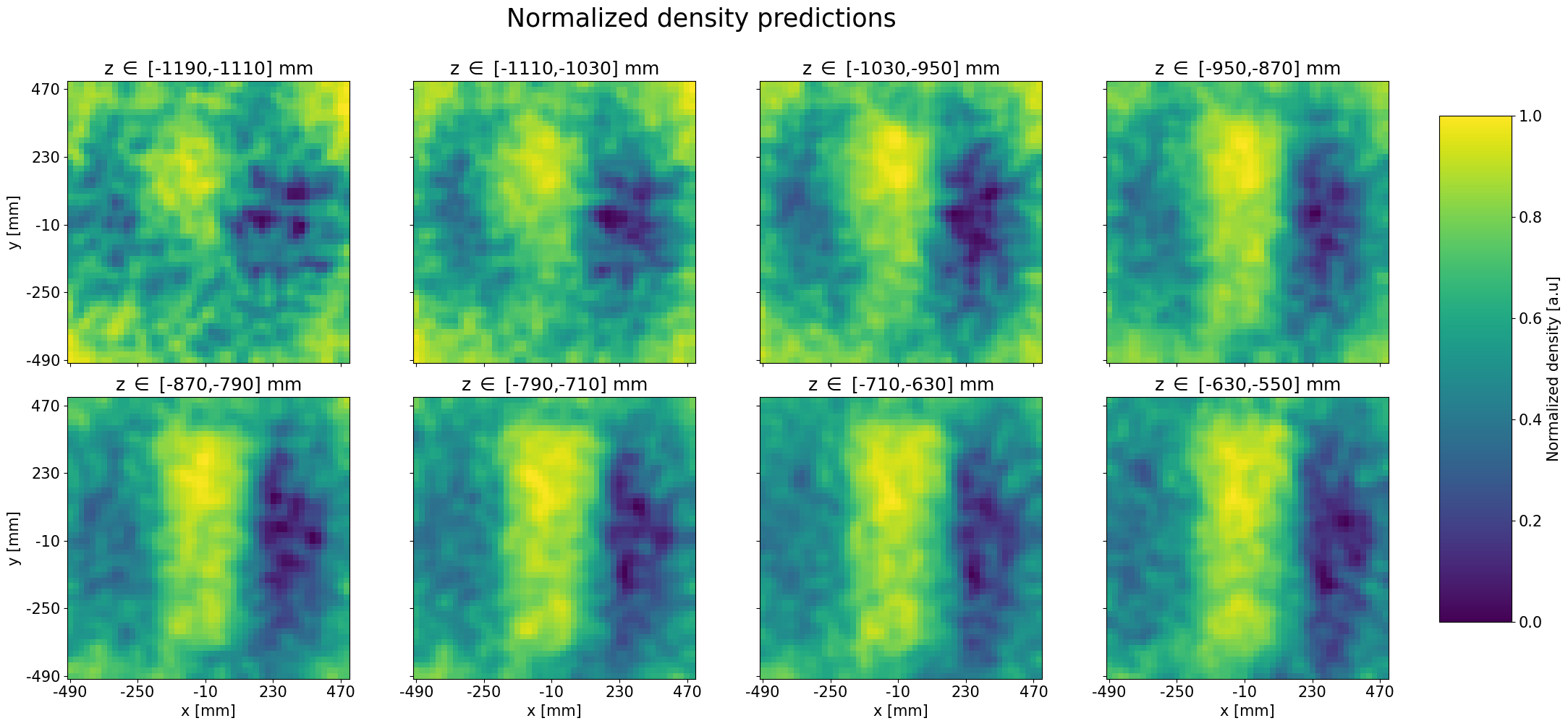}
    \caption{Normalized density predictions for the absorption technique using a back-projection algorithm. Predictions are shown by slice of 3 voxels layers along z (vertical axis).}
    \label{fig:scale_16_abs_slices}
\end{figure}

\begin{figure}
    \centering
    \includegraphics[width=.70\linewidth]{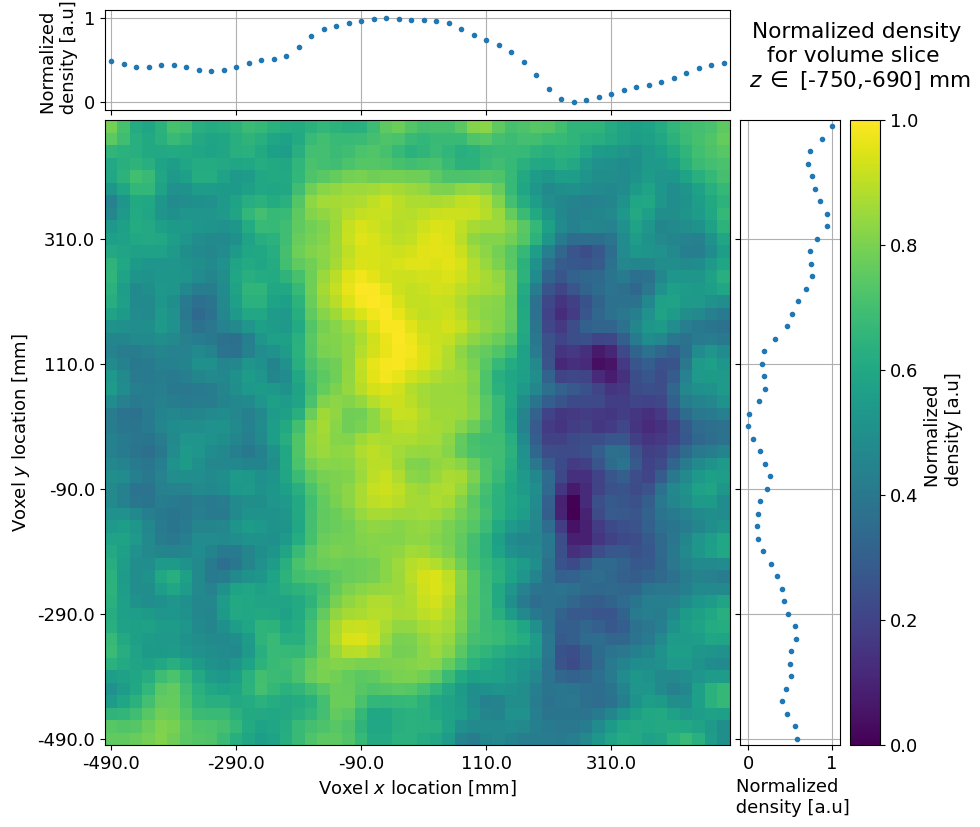}
    \caption{Normalized density predictions for the absorption technique using a back-projection algorithm. Predictions are shown for a slice of 3 voxels layers along z (vertical axis). The true location of the 40 cm diameter cylinder is $x, y, z, = (0, 0, -710) \text{mm}$.}
    \label{fig:scale_16_abs}
\end{figure}

Density predictions presented in Figure~\ref{fig:scale_16_abs_slices} and \ref{fig:scale_16_abs} are obtained with the same back-projection algorithm as in Section~\ref{sec:scenario_ii}, and smoothed with the same Gaussian smoothing algorithm.

Eight hours of data acquisition is sufficient to identify the presence of the cylinder within the statue. The image resolution along the $x$ and $y$ directions is good enough to evaluate the radius of the cylinder. Using the normalized density along the $x$ axis on Figure~\ref{fig:scale_16_abs}, the estimated cylinder diameter ranges between $34$ and $44\:\text{cm}$ which is compatible with the actual diameter value of $40\:\text{cm}$. However, the image resolution along the vertical $z$ axis is quite poor and does not allow to precisely evaluate the $z$ position of the cylinder, as shown in Figure~\ref{fig:scale_16_abs_slices}. Nevertheless, in the context of an actual measurement it is possible to combine measurements from different points of view. In this simulation, the statue is laid down horizontally, which is often unrealistic. In a real life scenario, one can advantageously orient the detection system towards the target and proceed to several measurements exploiting the azimuthal symmetry of the cosmic muon flux. These various measurement can then be combined into a single 3D density map, as presented in \cite{miyamoto2022tomographic}. 


\section{Proposal for future case studies}
\label{sec:future-studies}



This section presents three interesting applications in cultural heritage where muography can make significant contributions: estimates of the moisture content, imaging of hidden features, and detection of cracks. 
We present possible case studies to evidence the strengths and limitations of the method, and to motivate technical or methodological improvements and needs.

\subsection{Moisture content measurements}
\label{sec:moisture}

The ability to estimate the moisture content of sculptures or structures is important for (preventive) conservation and treatment proposals, as water is one of the main drivers for decay mechanisms such as freeze-thaw action, salt crystallization,  biological growth, hygric / hydric dilation, corrosion, and reduction of strength. 
Sources of moisture include (wind-driven) rain, rising damp, infiltration by leakage or flooding, and condensation.

The water content, expressed as a fraction of mass (or weight percent, abbreviated in wt-\%), can vary between 0 wt-\% in perfectly dry conditions and 30 wt-\% for various saturated porous materials (like natural stone, ceramics, mortar, wood). 
Typical moisture contents are in the range between 8 and 20 wt-\%.

\begin{figure}[!ht]
\centering
    \includegraphics[height=0.25\paperheight]{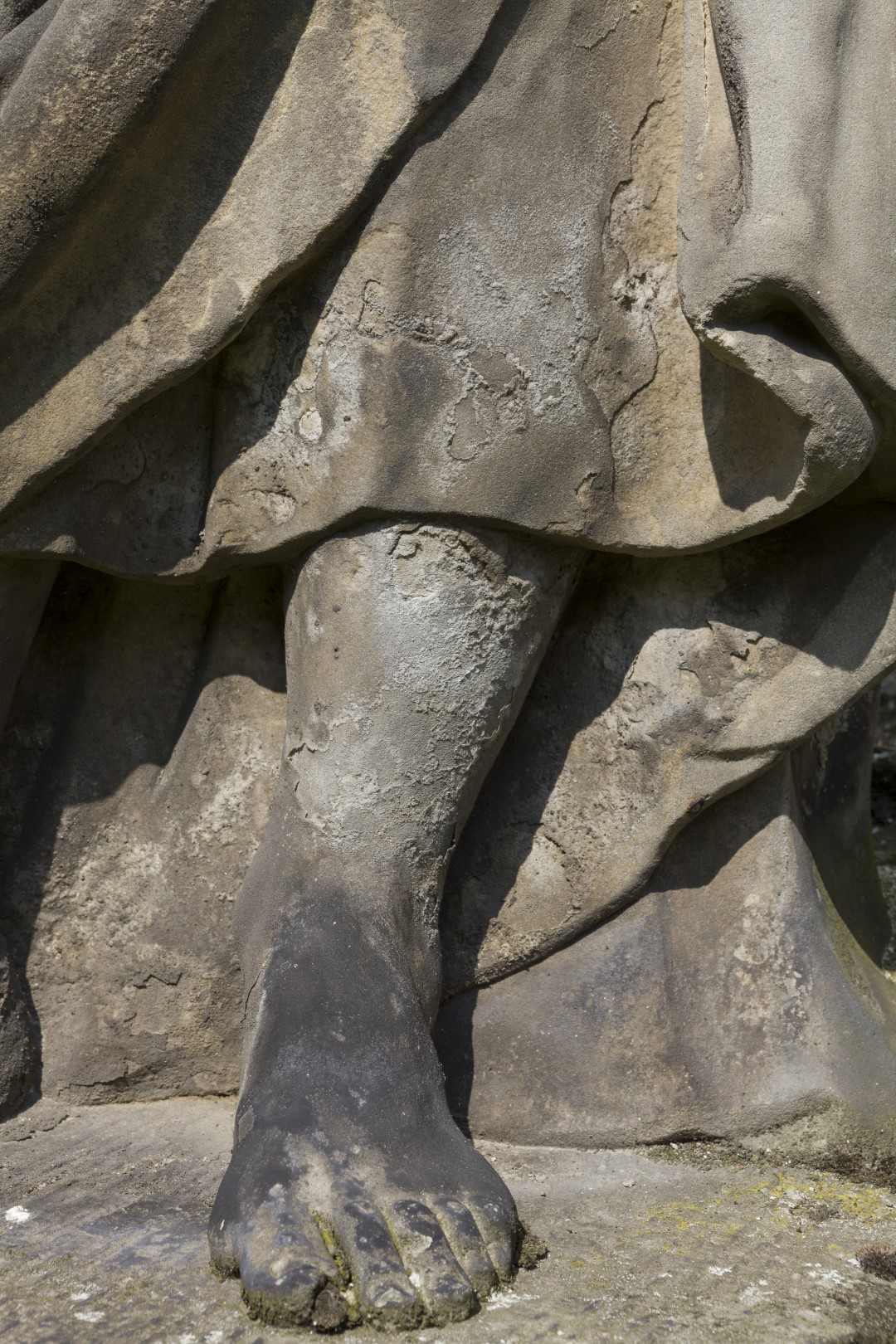}
    \caption{A detail of a statue with moisture and salt damage from the Calvary of St. Paul's church in Antwerp. Picture is \copyright KIK-IRPA, Brussels.}
 		\label{fig:moisture}
\end{figure}

Moisture measurements can be divided in direct and indirect measurements. Direct measurements, which are the most accurate, require {\it in situ} sampling: the artifact is drilled to collect powder for analysis. Quantitative measurements include gravimetric weighing and drying (i.e. weighing the sample before and after drying it). For stones, a quantitative assessment can be done by the calcium carbide test, in which the moisture content of the powder from drilling is determined based on the chemical reaction $H_2 O + CaC_2 \to CaO + C_2 H_2$ where the production of acetylene ($C_2 H_2$) gas increases the pressure in a closed container. 
The main drawback of sampling-based techniques is their destructive nature; moreover, the calcium carbide test may underestimate the moisture content because of incomplete chemical reactions.

Indirect measurements of moisture content are based on electrical resistivity, electrical capacity, microwaves, radar, thermal imaging, equilibrium methods and time-domain reflectometry~\cite{Orr2021HES}.
Their main strength is that they are non-destructive. However, they face several complexities; most probe only shallow depths (of order mm to cm), some are affected by the salt content and all are dependent on the contact between probe and material. Additionally, these techniques are generally considered as merely indicative of the actual moisture content, and quantification needs time-consuming calibrations per material ~\cite{Orr2019,Orr2020}. 

In this context muography can find a strong use case, being an {\it in situ} but non-destructive remote-sensing technique that reveals in-depth information as the muon passes through the object. 

Muography can measure moisture content thanks to its effect on density. 
The average density of the affected volume increases when wet, i.e. pores that would normally contain air would fill with water.
In cases where the volume is not altered and only mass is changing (as in the case of stones), adding moisture to a stone object of density $\rho$ results in a new density $\rho^\prime$ according to:
\begin{equation}
\rho = \frac{M}{V} \rightarrow \rho^\prime = \frac{M^\prime}{V} = \frac{M + M_{water}}{V} = \rho + \frac{M_{water}}{V} \, .
\label{eq:density-moisture}
\end{equation}

The moisture content (MC in wt-\%) is expressed as the mass-fraction of water with respect to the dry material mass, i.e. $M_{water} = {\rm MC} \times M_{dry}$, which substituted in Eq.~\ref{eq:density-moisture} yields $\rho^\prime = (1+{\rm MC})\times\rho$.  
The density of most stone fabrics is around 1.5 - 2.7 $g/cm^3$. If, for example, $\rho_{dry} = 2~g/cm^3$ and MC = 20 wt-\%, the corresponding water-saturated density is $\rho_{saturated} = 2.4~g/cm^3$. 
In another relevant example for cultural heritage preservation, we may have a density of $\rho_{healthy} = 0.8~g/cm^3$ for healthy construction wood at normal moisture levels (around 12\%), increasing to $\rho_{unhealthy} = 0.94~g/cm^3$ if a moisture pathology brings the moisture content up to 30 wt-\%~\footnote{This value corresponds to the average fibre saturation point of wood, at which in theory the decay will start. In practice, staining or mould growth can happen also at lower MC. In fact, decay kicks off when a liquid water source is available to the fungus, and some fungi are capable of transporting water from a distance.}. 
Such a  $\rho$ difference is potentially within reach for muography, and the difference in muon flux attenuation can be significant also when the difference in density due to moisture is not so large. 

The moisture content is dynamic, following wetting and drying, and it is a suitable problem for long-term (seasonal) muography monitoring, meaning that the observed muon flux through the monument would be plotted as a function of time. 
From the perspective of cultural heritage conservation, the most severe issues occur with permanent high moisture content, implying that even infrequent muography measurements would yield valuable information.
A relevant case study is offered by the 68 sandstone sculptures of the Calvary of the St. Paul's Church in Antwerp (Belgium), several of which suffer various forms of degradation attributed to the action of moisture~\cite{DeRoy2020}, as shown in Figure~\ref{fig:moisture}.

It can be noted that spatial resolution is not an important consideration for this particular use case, differently from most other use cases discussed in this paper, as moisture spreads broadly. Moreover, to some extent it is not important to have 3D imaging, as 2D projections can already be very informative. 
On the other hand, precision is requested on the flux measurement, as it gets translated into an opacity measurement (where opacity indicates the integral of $\rho$ along a line of sight)~\cite{Bonechi:2019ckl}, which is in turn used to estimate the moisture load and, if sufficient data are taken, its dependence on time.

\subsection{Bulk features in monuments and artwork}
\label{sec:inner-features}

\begin{figure}[!ht]
	\begin{center}
		\begin{subfigure}[t]{\sfSmall\textwidth}
 			\begin{center}
 				\includegraphics[height=\sfTiny\textheight]{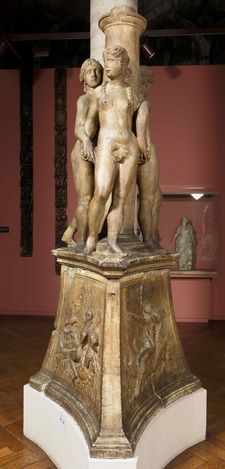}
 			\end{center}
 		\end{subfigure}
 		\begin{subfigure}[t]{\sfSmall\textwidth}
 			\begin{center}
 				\includegraphics[height=\sfTiny\textheight]{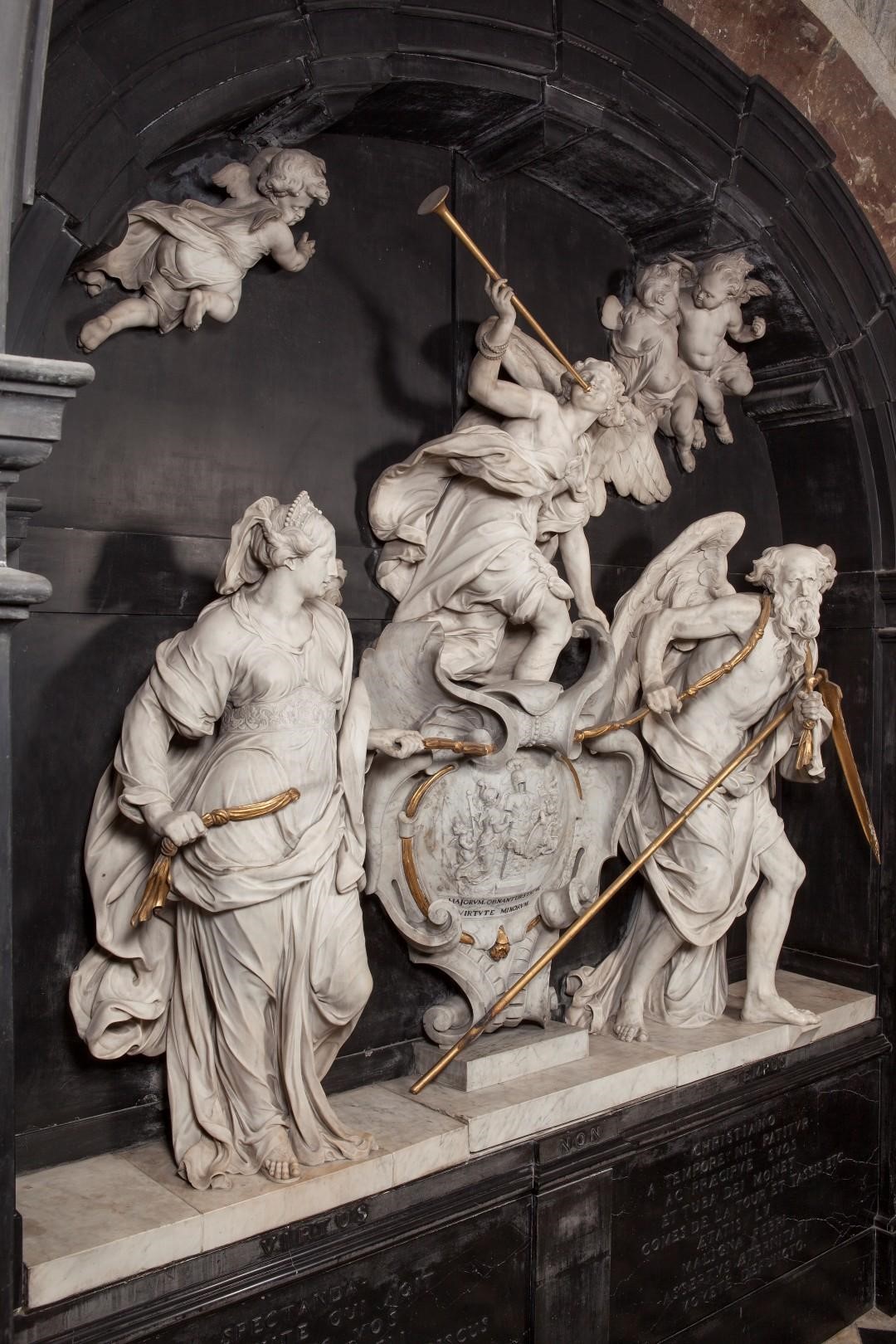}
 			\end{center}
 		\end{subfigure}
 		\caption{Left: The Fountain of the Three Graces, in the Museum of the City of Brussels. Right: The monument of Lamoral Claude François von Thurn und Taxis, by the sculptor Mattheus Van Beveren. Pictures are \copyright KIK-IRPA, Brussels.}
 		\label{fig:examples1}
 	\end{center}
\end{figure}

Bulk or inner features in monuments and artwork are typically not directly observable from the surface of objects. However, these contain crucial information on construction technology, interventions and material condition. Therefore, their identification is an essential, but not always evident part of the condition assessment of objects. 
Direct assessment of the bulk features requires dismantling of the structure, which is often not possible or at least risky, or destructive drill coring. Therefore, indirect and portable methods are valuable alternatives.

Muography offers the possibility to visualise inner features of a large and dense object, thanks to the high penetrating power of muons. 
This is particularly appealing for cultural heritage objects which contain or are suspected to contain non-trivial features that are invisible from the exterior. 
In this vast category of use cases, differently from the moisture measurement described previously, precision is more sought on the contrast ($\Delta \rho$ as function of position) than on the absolute value of $\rho$. 
Moreover, 3D information is usually important, demanding either more than one detector taking data simultaneously or the same detector taking data sequentially from more than one viewpoint. 

This is further illustrated by a few representative examples from Brussels (Belgium), chosen among those on which some of the authors have a direct and deep knowledge. 
All these examples are artwork made of dense stone materials, which are difficult or risky to move and because of their size not easy to image using more traditional imaging techniques. On the other hand their size is relatively small by the standards of muography, which means that imaging can be challenging. 

\vspace{1em}
The \textit{Fountain of the Three Graces} (Fig.~\ref{fig:examples1}, left), in the permanent collection of the Museum of the City of Brussels, was sculpted in white Carrara marble in the 1530's~\cite{patigny2021count}. 
The fountain is composed of two main volumes of white marble: the three Graces with the central column, and the triangular base with mythological scenes. We have no information about the inner tube system that is probably still present in the core center of the column and most likely connected to the nozzles in the breasts of the female figures and the nozzle in one of the mythological scenes on the base. 



The \textit{monument of Lamoral Claudius Franz von Thurn und Taxis} (Fig.~\ref{fig:examples1}, right), located in the Church of Our Blessed Lady of the Sablon (Brussels), was sculpted by Mattheus Van Beveren in 1678 in white marble with black marble cladding~\cite{thurn-und-taxis}. 
The structure with black marble cladding underneath the sculptures is deformed due to subsidence, and demands further investigation of the inner brick structure to evaluate its condition. As dismantling of the sculptures poses risks for their preservation, non-destructive imaging of the structure would help in the assessment. 

\begin{figure}[ht]
	\begin{center}
		\begin{subfigure}[t]{\sfSmall\textwidth}
 			\begin{center}
 				\includegraphics[height=\sfVeryTiny\textheight]{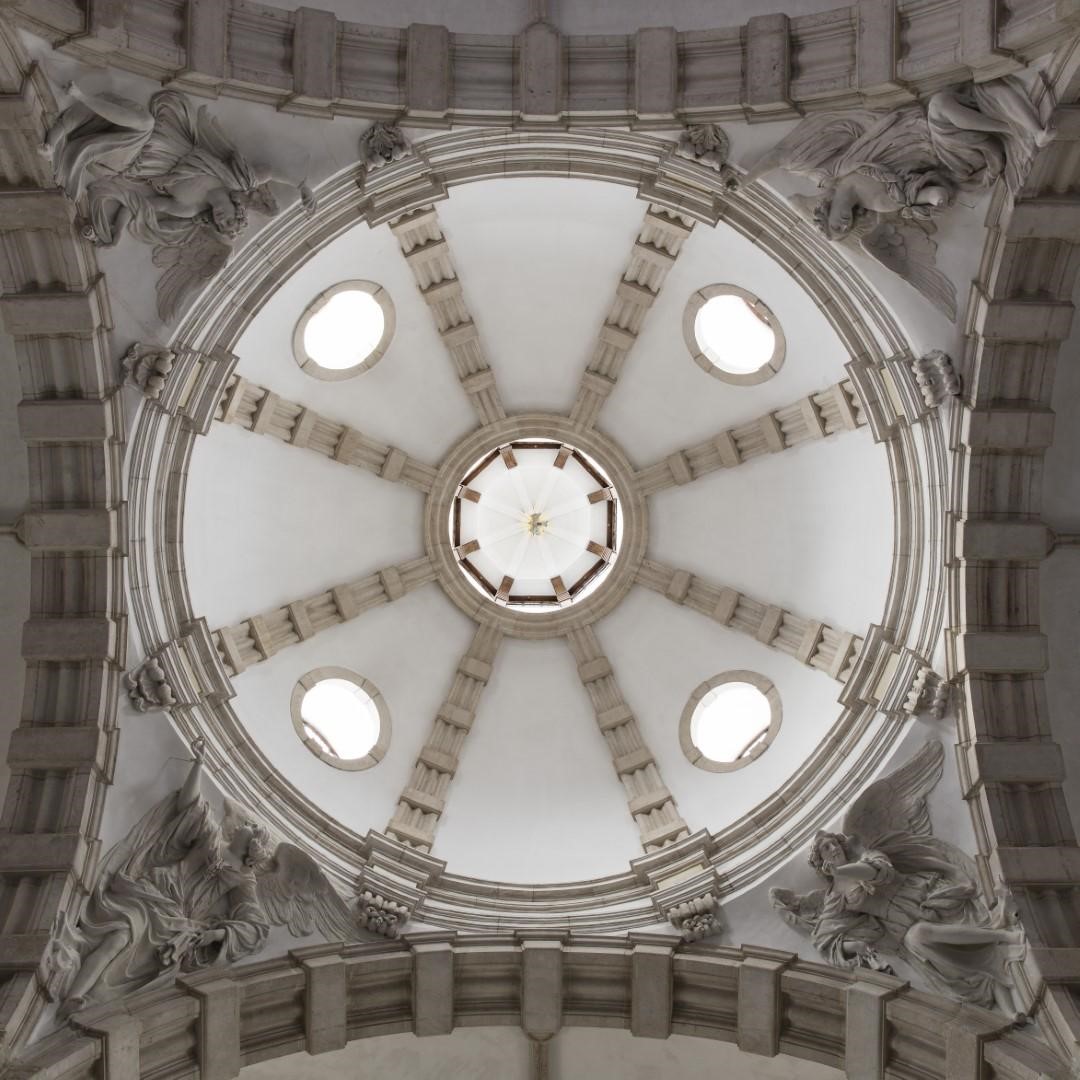}
 			\end{center}
 		\end{subfigure}
 		\begin{subfigure}[t]{\sfSmall\textwidth}
 			\begin{center}
 				\includegraphics[height=\sfVeryTiny\textheight]{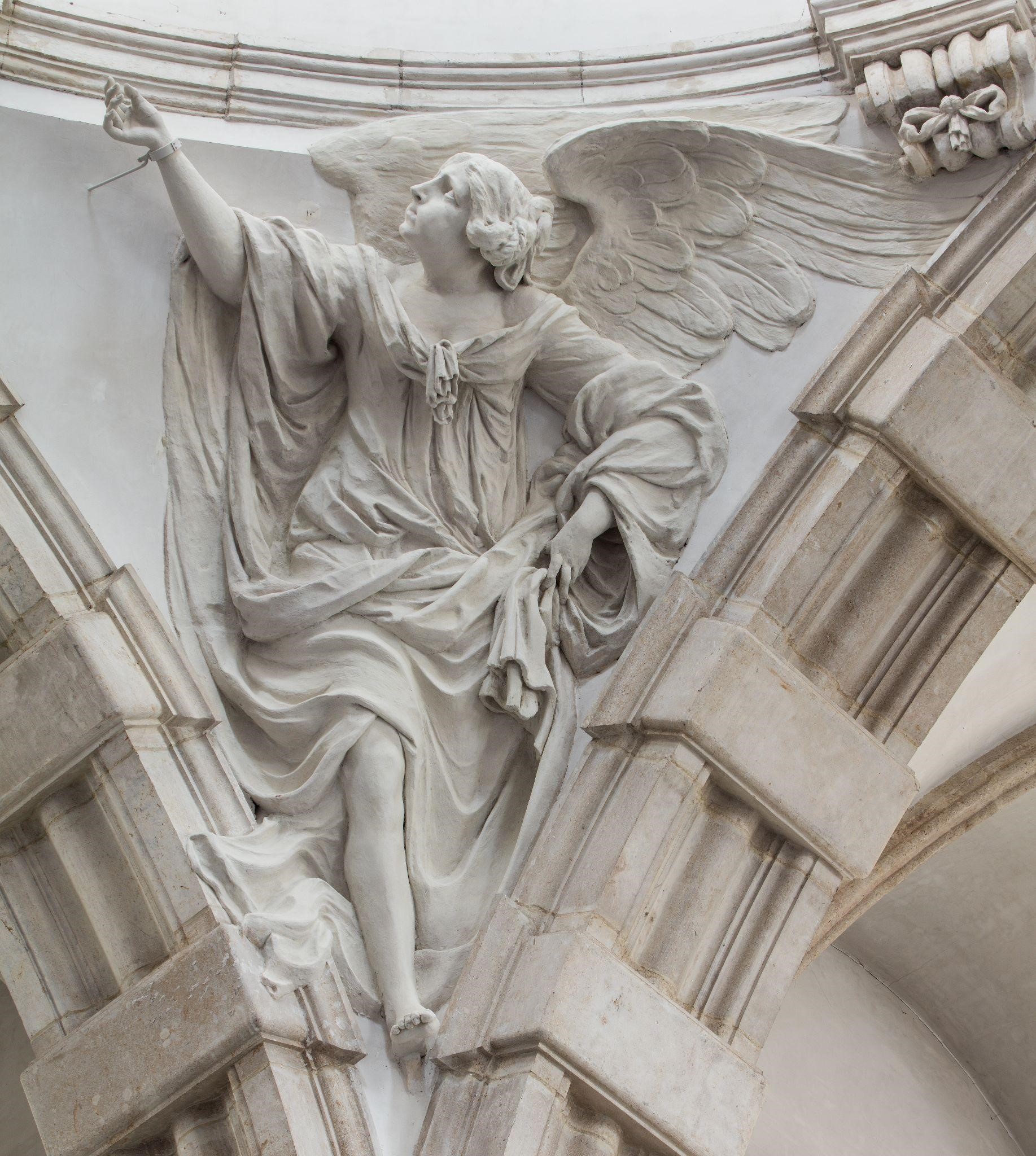}
 			\end{center}
 		\end{subfigure}
 		\caption{Left: Four sculpted angels in the pendants of the \'Eglise Notre-Dame aux Riches Claires. Right: close-up of one of the angels. Pictures are \copyright KIK-IRPA, Brussels.}
 		\label{fig:riches-claires}
 	\end{center}
\end{figure}

The pendants of the \textit{Notre-Dame aux Riches Claires} church in Brussels (Fig.~\ref{fig:riches-claires}) host four sculpted angels, composed of different volumes of Avesnes limestone, fixed in masonry of the pendants underneath the central dome of the church~\cite{riches-claires}. 
After a devastating fire in 1989, the sculptures were thoroughly restored. 
Unfortunately, these interventions were not documented in detail.  
In 2020, after the left arm of one of the angels fell down, a survey discovered several new cracks and fractures on all the four angels. 
Archival investigation of the work reports from the 1990's found mention of injections with a cement based mortar and pinning of fragments by iron threaded bars/dowels to the underlying structure of the vaults, but not their exact number and positions, thus impeding to assess if the current cracks and bad condition have anything to do with those interventions, which is an extremely important question from the point of view of cultural heritage restoration~\cite{HuysmansDeRoy2020}. 
Besides that, the angels were composed of different limestone blocks. The joints are slightly visible on the surface but it was not very clear how many blocks were used (some parts are made out of lime mortar as well and during restoration, new volumes were added with a mixture of plaster and stone dust). Moreover, the original anchors and the new added dowels would be interesting to localize.



\begin{figure}[ht]
	\begin{center}
		\begin{subfigure}[t]{\sfSmall\textwidth}
 			\begin{center}
 				\includegraphics[height=\sfTiny\textheight]{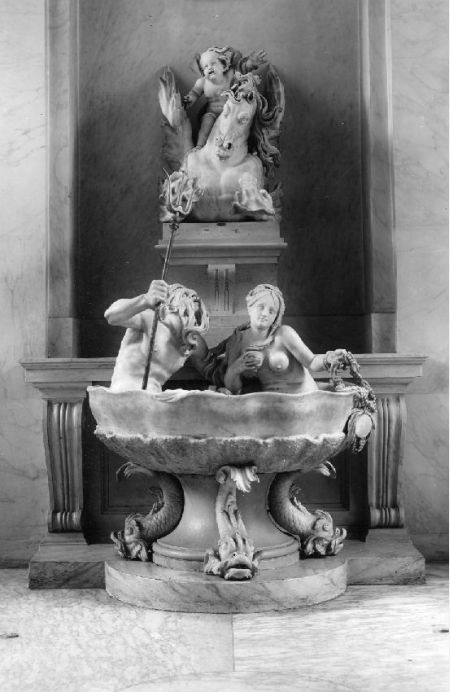}
 			\end{center}
 		\end{subfigure}
 		\begin{subfigure}[t]{\sfSmall\textwidth}
 			\begin{center}
 				\includegraphics[height=\sfTiny\textheight]{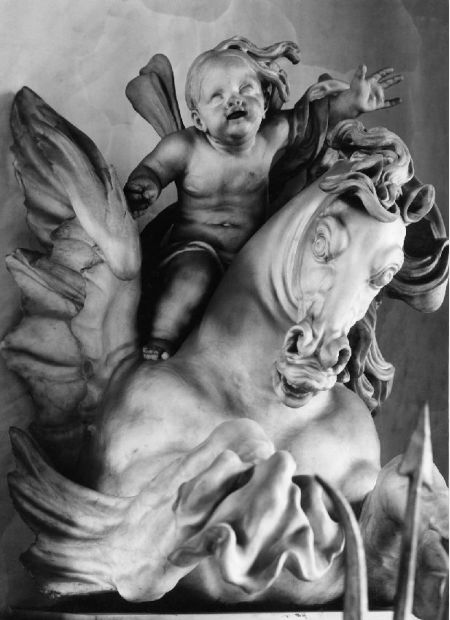}
 			\end{center}
 		\end{subfigure}
 		\caption{Left: Fountain of the sea gods, at the Museum of Fine Arts in Brussels. Right: close-up of one of the elements of the fountain. Pictures are \copyright KIK-IRPA, Brussels.}
 		\label{fig:sea-gods}
 	\end{center}
\end{figure}

The \textit{Fountain of the sea gods} (Fig.~\ref{fig:sea-gods}) was sculpted in white marble by Gabri\"el Grupello in 1676~\cite{sea-gods} and it is currently displayed in front of a staircase in the Museum of Fine Arts of Brussels. 
The fountain has an internal structure that would undoubtedly be interesting to investigate in view of its conservation treatment. As a matter of fact, traces of corrosion are visible on the surface of some parts of the fountain. 
The fountain can not be moved, making thus a case for portable muography detectors.



\subsection{Inner cracks}
\label{sec:inner-cracks}

Imaging a statue or another cultural heritage item may also be motivated by searching for cracks or other forms of defects that may appear due to material degradation. 
This can be challenging for muography, as in practice the crack would appear as a narrow zone of lower density. Therefore, for this goal, 3D spatial resolution must be pushed to the limit.

\begin{figure}[ht]
	\begin{center}
		\begin{subfigure}[t]{\sfSmall\textwidth}
 			\begin{center}
 				\includegraphics[height=\sfTiny\textheight]{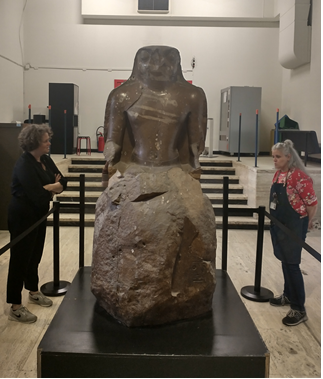}
 			\end{center}
 		\end{subfigure}
 		\begin{subfigure}[t]{\sfSmall\textwidth}
 			\begin{center}
 				\includegraphics[height=\sfTiny\textheight]{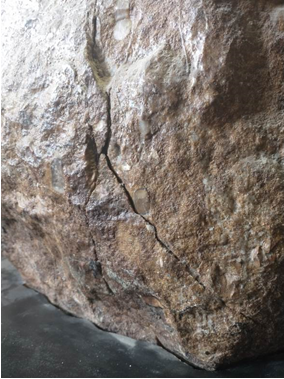}
 			\end{center}
 		\end{subfigure}
 		\caption{Left: Sculpture of the Egyptian deity Khonsou, at the Art \& History Museum of Brussels. Right: a close-up of one of the visible cracks. Pictures are \copyright KIK-IRPA, Brussels.}
 		\label{fig:khonsou}
 	\end{center}
\end{figure}

As an example, Fig.~\ref{fig:khonsou} shows a quartzite sculpture of the Egyptian deity Khonsou, dating from the 21st dynasty (1050 BCE), at the Art \& History Museum in Brussels~\cite{khonsou}.
This sculpture was seriously damaged due to a fire in the museum in 1946. Thereafter the sculpture was stored under an open shelter where it was exposed to frost and bad weather for a long time, causing many cracks and fractures.    
Due to its weight (estimated around 2500~kg) and its fragility, it is not recommended to move this object in this state of preservation. Yet the museum wants to move the sculpture for renovation works and a redesign of the collection. 
A well defined image of the inner fractures is necessary to evaluate its condition before conservation treatment and/or transport.  


Muography is recently starting to be considered as a promising non-destructive technique for searching for cracks, in the context of civil engineering, motivated by the monitoring of aging infrastructure (e.g. bridges). A proof-of-principle experiment has been reported in Ref.~\cite{niederleithinger2021muon}, based on muon scattering. 
For the same purpose, transmission-based muography has also been investigated~\cite{IAEA2022}. 
For a non-transportable item of the size of the statue described above, the latter appears to be more indicated, although the former is found to achieve a better spatial resolution.



\subsection{Summary of proposed case studies}

Table~\ref{tab:summary-case-studies-proposed} summarizes the muography requirements for the typologies of case studies proposed in this section. 
Although at this stage these considerations are purely qualitative, they can help guiding the strategy for detector development and deployment. 

\begin{table}
    \centering
    \begin{tabular}{c|c|c|c|c}
    Objective & Absolute $\rho$ & $\rho$ vs position & $\rho$ vs time & Material id. \\ \hline
    Moisture content & Important     & Not important & High frequency & Not needed \\
    Bulk features    & May be useful & Important & Not needed & May be useful \\
    Inner cracks     & Not needed & Very important & Low frequency & Not needed \\
    \end{tabular}
    \caption{Muography requirements for three categories of cultural heritage applications.}
    \label{tab:summary-case-studies-proposed}
\end{table}

When absolute density measurements are important, as in the case of estimating the wt-\% of water, it is necessary to also collect an auxiliary sample of muons under the same conditions but without the object of interest, in order to extract the transmission fraction from the ratio of the two datasets. 
However, in most cultural heritage use cases this is not crucial, as one is more interested in spatial resolution (hence $\rho$ as a function of position) or stability (hence $\rho$ as a function of time). In those cases, $\rho$ only needs to be expressed as a difference or a ratio with respect to a reference. In all cases, contrast between different $\rho$ values is important.

Spatial resolution in the measurement of $\rho$ vs position is always somewhat important when the feature being sought is a cavity or a crack, or a region of different material. 
Small features, including cracks, are challenging for absorption muography, therefore scattering muography with its superior intrinsic resolution should be preferred whenever it is realistic for the object to be sandwiched between muon detectors. 
Scattering muography is also to be preferred to absorption when the different material of the inner feature happens to be similar in density to the bulk material. 

Moisture content and inner cracks are two examples where it may be important to monitor the evolution of the object with time, in order to intervene at the first indications of a dangerous degradation of the conditions. 
The water fraction of an object can change quickly with time (especially outdoors), justifying in some cases a continuous monitoring, while the monitoring of cracks may be performed with infrequent data-taking campaigns. 


\section{Conclusions}
\label{sec:conclusions}

Muography is a promising tool due to its relatively low cost and portability. Thanks to the large muon penetration power, it is complementary to other imaging methods. 
We argue for the potential interest of muography as a nondestructive subsurface imaging technique in several use cases in cultural heritage conservation, where the size of the object of investigation is too large for other techniques. 
Muography has already been successfully employed for some very large cultural heritage objects such as pyramids, while its application to smaller objects has been largely unexplored.

In this paper, we reported a simulation study in ideal conditions with both scattering and absorption muography, assessing the strengths and limitations of both techniques for statues with different size.
Absorption and scattering have complementary strengths and weaknesses, but some limitations are in common for both: long acquisition times are necessary, due to the relatively low natural rate of cosmogenic muons, and muon direction and energy cannot be controlled.

In spite of those limitations, both techniques can provide unique data in various use cases, which we have qualitatively described with specific examples from our local context, which may become concrete targets for future muography studies.  
Targeting density differences between different inner parts of the object of interest, or searching for cracks, are typical use cases where the portability of muography detectors is appealing when the object of study is too fragile for transportation. Human-sized statues, however, appear to be at the lower limit of sensitivity of muography, motivating significant development work on precise instrumentation and cutting-edge algorithms for image reconstruction.
Moisture load measurement seems to be a ``low-hanging fruit'' for muography. At the best of our knowledge, this has not been exploited yet in the relevant literature.

\section*{Acknowledgments}
We thank KIK-IRPA for the figures in Section~\ref{sec:future-studies}. 
UGent-Woodlab and the Royal Museum for Central Africa (Cultural Anthropology and History Department) are acknowledged for providing Figure~\ref{fig:african-statue} and micro-CT data of that statue in the scope of the ToCoWo project (\url{https://tocowo.ugent.be/}). The ToCoWo project is funded by the Belgian Science Policy Office (BELSPO; B2/191/p2/TOCOWO). 
Jan Van den Bulcke of UGent-Woodlab provided helpful feedback on our considerations about estimating moisture content in wood.

A preliminary version of the simulation study reported in Section~\ref{sec:examples} has been presented at the Muon4Future workshop and in the corresponding proceedings~\cite{Muon4future}. We are grateful to several participants at that workshop for their constructive feedback.

This work was partially supported by the Fonds de la Recherche Scientifique - FNRS under Grants No. T.0099.19 and J.0070.21, and by the EU Horizon 2020 Research and Innovation Programme under the Marie Sklodowska-Curie Grant Agreement No. 822185.

\FloatBarrier
\clearpage
\appendixpage

\appendix\section{Algorithms}
\label{sec:appendix}

\begin{figure}[H]
    \centering
    \includegraphics[width=.3\textwidth]{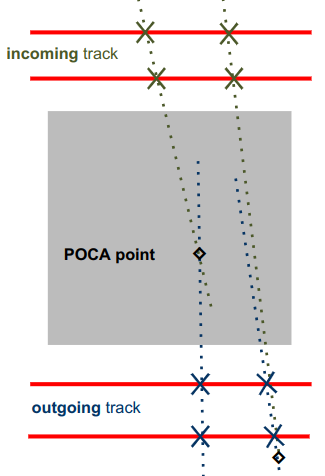}
    \caption{Reconstructed POCA point inside (left) and outside (right) the volume of interest.}
    \label{fig:poca_point}
\end{figure}

\paragraph{Point Of Closest Approach}
The Point Of Closest Approach algorithm works by finding the closest point between the incoming and outgoing tracks. It interprets this point as where the muon likely had a single high-energy elastic interaction with a nucleus. This approach neglects actual electromagnetic interactions that occurs along the its trajectory. Even though it is a rough approximation of reality, it has proven to be effective in many applications. Specifically, when the muon's scattering angle is large with respect to the angular resolution of the detection system, provides a good approximation for the muon scattering vertex. However, it has limitations when dealing with small scattering angles and nearly parallel incoming and outgoing tracks. Consequently, this leads to the reconstruction of POCA points outside the Volume of Interest (VOI), as illustrated in Figure~\ref{fig:poca_point}. In situations where spatial resolution is not perfect, as much as 40$\%$ of the POCA points can be reconstructed outside of the VOI. The problem can be somewhat resolved using an iterative POCA approach~\cite{kanth2014muon}.

The output of the POCA algorithm is a series of 4-dimensional data points; the $x, y, z$ locations as well as the scattering angle $d\theta$ between the incoming and outgoing tracks. Their spatial distribution reflect the density of the material and can be used as input of more advanced density reconstruction and clustering algorithms.

\paragraph{DBSCAN}

The DBSCAN algorithm \cite{DBScan} is one of the most commonly used clustering algorithm. It takes two parameters as input, $N_{\text{min}}$ and $\epsilon$, where $N_{\text{min}}$ is the minimum number of points within a radius $\epsilon$ (an ``$\epsilon$-neighborhood'') to form a dense region. Starting from a random point that has not been visited, the number of points $N$ within its $\epsilon$-neighborhood is computed. If $N >N_{\text{min}}$, a cluster is started, otherwise, the point is labeled as noise. If a point is found to be a dense part of a cluster, its $\epsilon$-neighborhood becomes part of that cluster. These steps are repeated until the whole density-connected cluster is found. Then, a new unvisited point is retrieved and processed, leading to the discovery of further clusters or noise.

When applied to POCA locations, the main limitation of this algorithm lies in finding the appropriate values of $\epsilon$ and $ N_{\text{min}}$. Given that the distribution of POCA locations is affected by various parameters such as exposure time, material density or even the position of the volume within the detector's acceptance, their is no such thing as an optimal parameters choice that would suit every muon scattering tomography experiment. In this study, $\epsilon$ and $ N_{\text{min}}$ were obtained through brute force parameter scanning.

\paragraph{Neighborhood sum}
The Neighborhood Sum algorithm~\cite{kumar2022comparative}, adeptly combines density of scattering points and scattering angle values.
This is achieved by computing the sum of scattering angles within a defined neighborhood, a circle in 2D or a sphere in 3D scenarios.
Notably, this method excels in material discrimination, leveraging the dependency of scattering points and angles on material density. Regions with high density points and scattering angles are accorded higher weight, while noise points, including those with elevated scattering angles, are effectively suppressed due to the lack of neighboring points.
This method performs well in situations with few scattered vertices, benefiting from efficient data aggregation to reduce reliance on individual points and enhance parameter estimation.

\paragraph{Binned Clustered Algorithm}

The Binned Clustered Algorithm was developed in \cite{BCA} is an extension of the POCA algorithm. It assigns to each voxel a score by taking into account the degree of spatial clustering of the POCA locations in addition to the associated scattering angle value. For each pair of POCA locations $i$, $j$ in a each voxel, a metric $s_{ij}$ is computed as:
$$
s_{ij} = \frac{d_{ij}}{\theta_i\theta_j} \, ,
$$
where $d_{ij}$, $\theta_i$, $\theta_j$ are respectively the euclidean distance between $i$ and $j$, the scattering angles associated to $i$ and $j$. Finally, the final score of each voxel is computed as the median of its $s_{ij}$ distribution. In order to reduce noise, the final score of voxels with less than five POCA locations is set to $0$.

\paragraph{Gaussian smoothing}

Gaussian smoothing is a commonly used technique in image processing, allowing to reduce the noise of an image. Mathematically, it corresponds to convolving the image with a Gaussian function. Each pixel value is modified to a weighted average of that pixel's neighborhood, with weights computed as a 2D Gaussian:
$$
\mathcal{G}(x, y) = \frac{1}{2\pi\sigma^2} e^{-\left( x^2+y^2\right) / 2\sigma^2} \, ,
$$ 
where $x, y$ are neighboring pixel positions with respect to the central one. 

\paragraph{Backprojection algorithm}

\begin{figure}
    \centering
    \includegraphics[width=.3\textwidth]{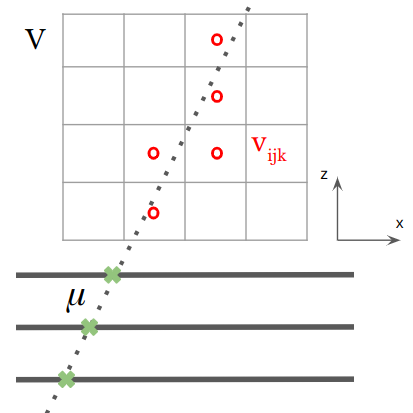}
    \caption{Back-projection of a reconstructed muon track into a voxelized volume of interest $V$ in the context of absorption-based muography experiment. The voxels $v_{ijk}$ traversed by the muon track are marked by a red circle.}
    \label{fig:vox_triggering}
\end{figure}

As discussed in Section~\ref{sec:state-of-the-art-muo}, most of state of the art absorption-based muography experiments are dealing with large objects, i.e from tens of meters to kilometers. The detection system being a few orders of magnitude smaller, and/or placed so far away from the target that it can be considered as a point-like object. In this context, one only measures the muons' direction $(\theta,\phi)$ and compute the transmission map as a 2D histogram with $\theta$ and $\phi$ bins. However, to study human-sized sculptures we have in general the possibility to position the detectors very close to the statue, in order to maximize the resolution within the object, and this approximation is no longer valid. One can use the muons' $(x,y)$ position as an additional measurement, making it possible possible to back-propagate muon tracks into the volume of interest and obtain density information about the volume based on the number transmitted muons. Such 3D reconstruction can be done using a fairly simple algorithm, inspired from \cite{ASR} and \cite{BonechiBP}, and presented in Figure~\ref{fig:vox_triggering}:

\begin{enumerate}
    \item Define a volume V within detector acceptance
    \item Divide the volume into voxels of size $d$.
    \item For each voxel, count the number muon tracks passing through it.
    \item At the end of the scan, each voxel has a number of muons $N$ associated to it.
\end{enumerate}

One needs two datasets: one with free sky and one with the target object. Then the muon transmission ratio of each voxel computed as $\frac{N_{\text{object}}}{N_{\text{freesky}}}$ can be used for estimating the object's average density.

\FloatBarrier
\clearpage
\small{\bibliography{refs}}

\end{document}